\documentclass[acmsmall,screen, preprint, 
nonacm,review=false,timestamp=false]{acmart}  
\setcopyright{none}
\AtBeginDocument{%
  }

\usepackage{tikz}
\usetikzlibrary{fit,backgrounds,arrows.meta,calc}
\usepackage{array}
\usepackage{multirow}
\usepackage[table,xcdraw]{xcolor}
\usepackage{ragged2e} 
\usepackage{booktabs}
\usepackage{enumitem}

\usepackage[table]{xcolor}
\usepackage{array}
\usepackage{ragged2e}
\usepackage{multirow}
\usepackage{booktabs}
\usepackage{tabularx}

\newcolumntype{L}[1]{>{\RaggedRight\arraybackslash}p{#1}}
\newcolumntype{G}[1]{>{\RaggedRight\arraybackslash\columncolor[HTML]{F3F3F3}}p{#1}}

\normalmarginpar

\begin{document}






\title[Chatbot Emotional Support Among People with Suicidal Ideation and other Mental Health Concerns]{\textit{“I Just Don’t Want to Be a Burden”:} Investigating AI Chatbot Emotional Support Among People with Suicidal Ideation and Other Mental Health Concerns}


\author{Vasudha Varadarajan}
\authornote{These authors contributed equally to this research.}
\affiliation{%
  \institution{Carnegie Mellon University}
  \city{Pittsburgh}
  \state{Pennsylvania}
  \country{USA}
}
\email{vvaradar@andrew.cmu.edu}
\orcid{0009-0004-0238-392X}
\author{Nathan S. Fishbein}
\authornotemark[1]
\email{nfishbein@fordham.edu}
\orcid{0000-0003-0769-4896}
\affiliation{%
  \institution{Fordham University}
  \city{New York}
  \state{NY}
  \country{USA}
}
\author{Andrew Aquilina}
\authornotemark[1]
\affiliation{%
  \institution{University of Pittsburgh}
  \city{Pittsburgh}
  \state{Pennsylvania}
  \country{USA}}
\orcid{0009-0002-8148-8353}
\email{andrew.aquilina@pitt.edu}

\author{Jordan Taylor}
\affiliation{%
  \institution{University of Michigan}
  \city{Ann Arbor}
  \state{Michigan}
  \country{USA}
}
\email{jordant@andrew.cmu.edu}
\orcid{0000-0002-0896-992X}

\author{Hong Shen}
\affiliation{%
 \institution{Carnegie Mellon University}
 \city{Pittsburgh}
 \state{Pennsylvania}
 \country{USA}}
 \orcid{0000-0002-5364-3718}

\author{Maarten Sap}
\affiliation{%
 \institution{Carnegie Mellon University}
 \city{Pittsburgh}
 \state{Pennsylvania}
 \country{USA}}
 \orcid{0000-0002-0701-4654}

\renewcommand{\shortauthors}{Varadarajan et al.}

\begin{abstract}
People experiencing high-acuity mental health concerns are turning to AI chatbots for emotional support, but little is known about what drives this use, their experiences, and their preferences. In a sequential mixed-methods study, we inductively analyzed Reddit posts and used the findings to develop a survey. We surveyed 202 adults with recent mental health concerns who had used AI for emotional support. Most participants were already receiving clinical care and reported turning to AI, most commonly because they did not want to burden others, rather than being driven by barriers to care. Benefits were broadly shared, but participants reporting past-year suicidal ideation described more harm-facilitating AI behavior and reliance than those without. Overall, participants ranked honesty above warmth, and preference for crisis escalation declined as clinical acuity rose. These findings motivate AI support that addresses perceived costs of disclosure, evaluates benefits alongside harms, and widens feasible paths to human connection.
\end{abstract}


\begin{teaserfigure}
\centering
 \includegraphics[width=0.8\textwidth]{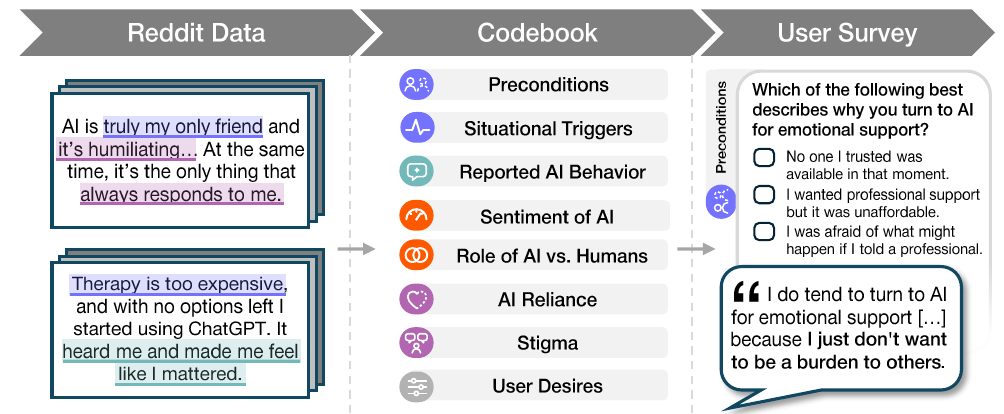}
  \caption{\textbf{Overview of the sequential mixed-methods study design.} We first analyzed Reddit posts describing AI use and mental health-related distress and inductively developed a codebook. We then translated these constructs into structured survey items to assess whether the patterns identified in Reddit appeared in a broader sample of people who had used AI for emotional support and reported psychiatric diagnoses or high-acuity mental health problems. Reddit excerpts shown in the figure are paraphrased.}
  \Description{Three columns are connected by arrows, pointing from left to right: Reddit Data, Codebook, and User Survey. In the Reddit Data column, there are two excerpts reading “AI is truly my only friend and it’s humiliating… At the same time, it’s the only thing that always responds to me” and “Therapy is too expensive, and with no options left I started using ChatGPT. It heard me and made me feel like I mattered.” Phrases that correspond to codebook categories are highlighted. In the Codebook column, the following categories are listed alongside small illustrations reflecting their meaning: preconditions, situational triggers, reported chatbot behavior, sentiment, role relative to humans, reliance, stigma, and user desires. The survey column includes an example survey question with checkboxes and a speech bubble with a quote from the survey reading “I do tend to turn to AI for emotional support [...] because I just don’t want to be a burden to others.”}
  \label{fig:teaser}
\end{teaserfigure}


\maketitle

\textbf{Content Warning: This paper discusses mental health topics that may be distressing to some readers.}

\section{Introduction}

In the United States, suicide rates rose 37\% from 2000 to 2018 and have remained near record highs since then \cite{garnett2022suicide,garnett2025suicide}. People with mental health problems, including suicidal ideation, often struggle to receive timely support. Formal care is difficult to reach: nearly half of adults with a mental illness did not receive treatment in 2024~\cite{briones2026mental}, and over 163 million Americans live in areas with a shortage of mental health providers~\cite{belz2024lessons}. Though informal support from friends, family, or peers can protect against psychological distress~\cite{cohen1985stress}, this type of support is increasingly rare, and the Surgeon General and WHO both recently declared loneliness a public health priority~\cite{WHOreport2025,officeofgeneralsurgeon2023}. Critically, how people perceive their relationships may shape both suicidal desire and their willingness to seek support. The interpersonal theory of suicide identifies perceived burdensomeness, or the belief that others would be better off without oneself, as a contributor to suicidal desire \cite{van2010interpersonal}. Prior work supports an association between perceived burdensomeness and suicidal ideation \cite{chu2017interpersonal}. These issues raise questions about where people can and do turn during acute distress and how their perceptions of their relationships shape whether available support feels possible to seek.

People are increasingly turning to generative AI chatbots for emotional support, using systems such as ChatGPT to disclose distress, seek advice, and work through personal difficulties~\cite{jung2025ve, siddals2024happened}.
Prior research suggests that these systems can feel accessible, private, and nonjudgmental, including for people navigating social anxiety~\cite{Wang_Wang_Crace_Zhang_2025}.
Beyond general-purpose chatbots, AI companions can become relational figures in users' lives, offering a sense of connection while also raising questions about dependency, well-being, and the appropriate role of AI in mental health support~\cite{Yuan_Zhang_Aledavood_Zhang_Saha_2026}.

Despite growing attention to generative AI in mental health, we know comparatively little about its use by people experiencing mental health concerns that may involve risks of serious harm, including suicidal ideation, delusions, and mania. 
Such use is not confined to specialized clinical tools: acute episodes unfold through the same general-purpose systems used for everyday, lower-stakes emotional support~\cite{luo2025seeking}. Although OpenAI estimates that only 0.15\% of weekly active ChatGPT users have conversations containing explicit indicators of potential suicidal planning or intent, this proportion represents a substantial population at platform scale~\cite{StrengtheningOpenAI2026}. 
Studies of chatlogs have documented how chatbots may amplify delusional beliefs when users experience suicidal ideation~\cite{moore2026characterizing,aquilina2026lost}.
Further, existing user studies evaluating support-seeking on social media do not distinguish between general emotional support and support for acute, high-risk mental health conditions~\cite{jung2025ve}. 

Existing LLM safety research has largely taken a complementary approach, in which benchmarks typically evaluate model responses against expert-defined safety criteria~\cite{belli2026vera, weilnhammer2026clinically}. 
These evaluations are critical, 
but they often treat crises as discrete and readily identifiable exchanges, potentially assuming that users recognize that they are in crisis, seek a clinical form of intervention, and can act on referrals to formal care.
User-centered studies have examined the social contexts that shape help-seeking, including the design and availability of mental health resources, as well as users' attitudes toward AI during crises -- often developed through in-depth interviews with smaller populations~\cite{yoo2025ai, pendse2021can, siddals2024happened}. 
We build on this foundation by examining the experience more holistically, seeking to develop a broader account of 
 \textbf{what leads people with psychiatric diagnoses or high-acuity mental health problems, including suicidal ideation, to use AI for support, how those exchanges unfold, and how users perceive and respond to them afterward}. Specifically, we ask:
\begin{itemize}
\item \textbf{RQ1}: What situations and circumstances lead people in this population to seek support from chatbots?
\item \textbf{RQ2}: How does this group describe their experiences of using chatbots for support?
\item \textbf{RQ3}: What do people in this group need and want chatbots to do, and not do, when providing emotional or crisis-related support?
\end{itemize}

To address these questions, we conducted a sequential mixed-methods study that uses exploratory qualitative analysis to identify patterns of AI use among those reporting mental health concerns and a subsequent survey to assess whether and how those patterns generalize across a broader population of users.

We first analyze Reddit discussions in which users describe mental health concerns and turning to generative AI for emotional support. Using a hybrid inductive-deductive analytic approach, we develop a codebook capturing the circumstances preceding AI use, the forms of support and interaction that emerge, and users' subsequent appraisals of AI. This exploratory analysis provides a detailed and user-described account of heterogeneous experiences of AI use among people reporting mental health concerns across Reddit communities, while generating empirically grounded patterns and questions for further examination.
We then use these findings to inform the design of a targeted user survey of Prolific participants who report lived experience of psychiatric diagnoses or high-acuity mental health problems, including suicidal ideation, and prior use of AI for support during distress. The survey assesses how patterns identified in the Reddit data resonate across a broader set of experiences and enables more direct examination of users' needs, expectations, current use, and changing relationships with AI across stages of seeking support. 

Drawing on both studies, we discuss implications for mental health practitioners, LLM developers, AI safety and trust teams, and platform-level safeguards for AI use among our target population, as well as opportunities to distinguish use during acute crises from lower-stakes forms of emotional support.
This work makes three contributions:

\begin{itemize}
    \item A situated account of AI-mediated help seeking, describing 
    AI use in the setting of self-reported psychiatric diagnoses or high-acuity mental health problems, connecting the circumstances that lead to use, the interactions that unfold, longer-term reliance, and users' desired and undesired AI behaviors.
    \item An empirical study of our account of AI use among this high-risk population, evaluated to parse the conditions, experiences and desires of participants with different clinical concerns, showing how interpersonal disclosure costs shape AI use, rather than access barriers alone, and how perceived benefits coexist with unevenly distributed costs 
    \item Implications for LLM developers and clinicians working with patients who use AI for emotional support.  We highlight the design tension between user agency and safety-oriented escalation, showing that preferences for safety intervention vary with users' proximity to risk.
\end{itemize}

\section{Related Work}




\subsection{Specialized Chatbots for Mental Healthcare}

The potential of conversational systems to expand access to mental health support has long been studied, with earlier systems designed for specific interventions. For instance, Woebot delivered cognitive behavioral therapy content in a randomized trial with young adults, while Wysa combined conversational support with structured exercises~\cite{fitzpatrick2017delivering,inkster2018empathy}. This literature found promising evidence but substantial variation in system design, study quality, and outcome measurement~\cite{gaffney2019conversational}. More recently, specialized systems have incorporated generative AI. Therabot, an expert-fine-tuned generative chatbot, was evaluated in a randomized trial for mental-health treatment~\cite{heinz2025randomized}. This work demonstrates the capacity of conversational agents to deliver psychological interventions under controlled conditions.

However, this research leaves an important form of support seeking unexamined: how and why people seek emotional support from general-purpose chatbots not designed as mental-health interventions. Mental health exchanges occur with chatbots used for everyday questions, writing, or other tasks, without users first selecting a dedicated mental health intervention~\cite{jung2025ve,song2025typing}. Studying use outside research settings is particularly important because self-directed digital mental health interventions show a substantial research-to-practice gap, with real-world engagement falling well below trial engagement~\cite{baumel2019trial}. Understanding these exchanges requires examining what makes a general-purpose system an accessible and acceptable place to disclose distress.

\subsection{General-Purpose Chatbots Used for Mental Healthcare}

The broad conversational appeal of general-purpose chatbots allows users to move between practical questions and emotional concerns within the same interface. For an existing user, emotional disclosure can therefore begin within a familiar tool, without a separate search for a mental-health service. Research describes users valuing immediate availability, flexible dialogue, and a perceived absence of judgment or interpersonal obligations~\cite{jung2025ve,siddals2024happened}. Some participants in \citet{siddals2024happened} also contrasted generative chatbots favorably with the scripted interactions of earlier mental-health apps, pointing to improved conversational flexibility and personal connection. These findings help explain how support-seeking can emerge through an accessible conversational tool.

Analyses of Reddit discussions describe chatbots providing validation, reflection, information, and concrete next steps, while also raising concerns about privacy, incorrect advice, and excessive agreement~\cite{jung2025ve}. Interview research likewise documents emotionally meaningful support alongside limitations in memory, guidance, and relational understanding~\cite{siddals2024happened}. To characterize these tensions, \citet{song2025typing} introduce \emph{therapeutic alignment}: aligning AI used for mental-health support with therapeutic values and users' broader needs. Their account emphasizes how users actively construct supportive roles for chatbots, including roles that complement human care. These needs are heterogeneous: people with differing levels of social anxiety place different weight on non-judgmental responding, emotional understanding, trust, and reliability~\cite{Wang_Wang_Crace_Zhang_2025}, while interviews with people living with mental-health conditions describe situational use and boundaries around what chatbots can offer during more complex distress~\cite{purohit2026conditional}.

This work situates AI use within both gaps in care and the perceived social costs of disclosure. Technology-mediated help-seeking research similarly shows that timing, availability, and the organization of services shape which forms of support people can use~\cite{pendse2021can}. However, the same accessibility that makes general-purpose chatbots appealing for everyday emotional support can also place them in roles with greater clinical stakes. The roles users assign to these systems may conflict with clinical recommendations, raising questions about how their responses should be evaluated in these moments.

\subsection{Clinical Recommendations for and Evaluations of Generative AI for Mental Health}

Normative research on generative AI and mental health has largely recommended that AI aid practitioners in providing care but not independently deliver treatment or function as a crisis resource. Some of this work proposes a staged model, in which AI advances slowly from simple tasks, such as psychoeducation, to more complex ones, such as treatment planning~\cite{stade2024large}. Autonomous care remains a distant, high-risk possibility~\cite{stade2024large}. This cautious approach reflects concerns that AI cannot assume the obligations and accountability of a therapist, particularly in risky interactions with suicidal users~\cite{khawaja2023your}. Some policymakers and professional groups have adopted similar perspectives~\cite{apaPsychiatric2026practice,illinois2025wopr}, and the American Psychological Association has called general-purpose chatbots ``unpredictable" in crisis management~\cite{apaPsychological2025advisory}. Despite these recommendations, many people continue to use general-purpose chatbots during acute distress~\cite{StrengtheningOpenAI2026}, leading experts to call for research assessing how these systems respond to mental-health concerns~\cite{hua2025charting}.

Though clinically grounded evaluations of general purpose chatbots are proliferating, none have been shown to predict safety or outcomes among high-risk real-world users. These evaluations typically examine model behavior under simulated conditions, using responses to vignettes or simulated users to compare AI behavior with clinician responses or against clinician-developed rubrics~\cite{mcbain2025competency,Moore_Grabb_Agnew_Klyman_Chancellor_Ong_Haber_2025,dwyer2025mindbench}. Performance varies by assessment: models respond appropriately to suicide-related questions at very high and very low risk levels but diverge from expert judgment at intermediate levels~\cite{mcbain2025alignment}; they judge the appropriateness of responses to suicidal ideation with varying accuracy, at times exceeding trained clinicians~\cite{mcbain2025competency}, and underperform clinicians in safely responding to content with delusions~\cite{Moore_Grabb_Agnew_Klyman_Chancellor_Ong_Haber_2025}. However, many assessments evaluate single-turn interactions and may therefore miss harms that accumulate across multi-turn conversations~\cite{morrin2026journey}. While benchmarks designed to evaluate risk over multiple turns are emerging~\cite{belli2026vera,bentley2026validation}, they remain simulated. Clinical trials provide another means of assessing outcomes related to chatbot behavior over multi-turn interactions. While they report reduced symptoms in mild-to-moderate clinical samples, they tend to exclude those with acute or high-risk conditions and use specialized AI~\cite{heinz2025randomized,shoshani2026efficacy}. Their applicability to general-purpose systems used during crises is therefore limited.

User-centered research provides complementary evidence about the circumstances surrounding use and how users perceive the support they receive. These studies use retrospective interviews, user accounts, and donated chat logs to study experiences with AI mental health support~\cite{moore2026characterizing, siddals2024happened, song2025typing, purohit2026conditional}. A small portion of these studies report on crisis use and delusional spirals~\cite{moore2026characterizing, ajmani2026seeking}. These studies report subjective experiences and interpretations that AI-behavior-focused research is unable to evaluate. However, the contexts that prompt people to seek emotional support from AI, how people describe and evaluate that support, and their preferences for AI behavior during distress remain incompletely connected across this literature. Our study examines these dimensions together among people with recent mental health concerns, including suicidal ideation.


\section{Study 1: Reddit study}
We first grounded our investigation in existing Reddit discussions to understand how people were already describing AI use for emotional support in their own terms.
Reddit's pseudonymous, discussion-oriented communities offer a relatively low-barrier setting for users to share perceived benefits as well as concerns, and recent AI-focused communities have produced substantial discourse around AI therapy, companionship, and emotional reliance~\cite{pataranutaporn2025my,aghakhani-rezapour-2026-like}. 

\label{sec:reddit-study}
\subsection{Data Collection}
\paragraph{Subreddit and Post Identification}
Inspired by \citet{shen2026ai}, we used an iterative, search-based procedure to identify Reddit communities where users discussed AI use alongside severe emotional distress or suicide-related concerns. We initially searched Reddit using \texttt{AI use}, \texttt{mental health}, \texttt{suicidal}, and \texttt{emotional support}, reviewed resulting posts and subreddits for relevance, and iteratively refined the query set using terminology from relevant results. The final terms were: \texttt{mental health, emotional support, loneliness, depression, psychosis, anxiety, therapy, AI use, AI for mental health, therapist, talk to AI, suicidal,} and \texttt{AI companion}.

We continued until thematic saturation, when additional searches identified no new relevant subreddits. This yielded 33 subreddits: \texttt{EDAnonymous}, \texttt{addiction}, \texttt{alcoholism}, \texttt{adhd}, \texttt{anxiety}, \texttt{autism}, \texttt{bipolarreddit}, \texttt{bpd}, \texttt{depression}, \texttt{healthanxiety}, \texttt{lonely}, \texttt{ptsd}, \texttt{schizophrenia}, \texttt{SuicideWatch}, \texttt{socialanxiety}, \texttt{mentalhealth}, \texttt{COVID19\_support}, \texttt{mentalillness}, \texttt{therapy},\\\texttt{therapists}, \texttt{raisedbynarcissists}, \texttt{askatherapist}, \texttt{BeyondtheAIAssistant}, \texttt{TherapyGPT}, \\\texttt{AI\_}\texttt{For\_}\texttt{Therapy}, \texttt{MyBoyfriendIsAI}, \texttt{AIRelationships}, \texttt{BeyondThePromptAI}, \texttt{ChatGPT}, \texttt{antiai}, \texttt{AITherapySpace}, \texttt{ClaudeAI}, and \texttt{CPTSD}. We removed duplicate subreddit names from the final list.

\paragraph{Data Collection and Filtering}We collected 3,342,335 posts published in these communities between January 1, 2023, and January 31, 2026, to capture AI use specifically in the post-LLM era. Sampling across communities captured diverse mental-health experiences, help-seeking contexts, and relationships with AI, rather than a narrower subset of Reddit users or experiences.

To identify posts discussing both AI use and elevated mental-health or suicide-related distress, we adapted the archetype-based suicide-risk inference method of \citet{varadarajan2024archetypes}. We curated separate prototype sets for (1) AI use and (2) severe emotional distress or suicide-related risk. Suicide-risk prototypes were adapted from \citet{varadarajan2024archetypes}, while AI-use prototypes were curated by the authors (see the Appendix~\S\ref{app:archetypes}).

We embedded prototypes and post sentences using \texttt{sentence-transformers/all-MiniLM-L6-v2}. For each construct, a post’s score was the maximum cosine similarity between any of its sentences and the relevant prototype set. Pre-specified thresholds of 0.30 for AI use and 0.50 for suicide-related risk yielded 2,008 posts meeting both criteria.\footnote{We selected thresholds by reviewing the 20 lowest-scoring posts retained under candidate cutoffs and choosing values for which at least 50\% of borderline posts mentioned the corresponding construct.}
The resulting candidate set contained posts referencing AI use in the context of mental-health concerns, including severe emotional distress and suicide-related risk.
\subsection{Codebook Development and Validation}

To characterize how Reddit users described their use of conversational AI in relation to mental-health support, 
we conducted an inductive qualitative analysis of 100 of the filtered Reddit posts. 
Rather than treating AI use as an isolated behavior, we examined it as situated within the posters' broader accounts of distress, available support, care access, interpersonal relationships, and everyday coping.
Our analytic focus was therefore the portion of each post in which the author mentioned using, or considering using an AI system, alongside the surrounding text needed to interpret the role that AI played in that account.

Two authors independently reviewed and conducted the analysis: one with training in clinical psychology and mental-health research, and one with expertise in computer science. 
This interdisciplinary pairing supported attention both to the psychosocial context of posters' narratives and to the particular affordances, risks, and interactional roles attributed to AI systems. 
We used a hybrid inductive and deductive approach~\cite{fereday2006demonstrating}, allowing codes to emerge from participants' descriptions rather than applying a pre-determined framework of mental-health needs, technology use, or AI harms.\looseness=-1

\paragraph{Inductive Analysis} We used a two-stage qualitative coding process. First, two authors independently open-coded the same 100 Reddit posts, focusing on passages in which posters described using, considering, or reflecting on AI and the surrounding context needed to interpret AI’s role. Researchers generated descriptive codes inductively, without a predetermined codebook. These codes captured explicit reasons for AI use (e.g., therapy cost or unavailability), as well as associated conditions, experiences, and perceived consequences, including immediate emotional release, feeling understood, frequent use, and concern about reliance on AI.\looseness=-1

The researchers then discussed candidate codes, and one synthesized an initial codebook by identifying recurring concepts, consolidating overlaps, and drafting definitions and inclusion criteria. 
The second researcher reviewed and revised this codebook, proposing additional codes and refinements to code boundaries. Disagreements were resolved through discussion, resulting in a shared codebook representing both researchers' interpretations of the initial 100 posts.\looseness=-1

\paragraph{Deductive Analysis} Using the shared codebook, each researcher independently coded an additional 30 posts, applying established codes while flagging unclear or insufficiently represented cases. Posts could receive multiple codes, reflecting intersecting motivations, constraints, use patterns, and perceived outcomes. For example, a poster might describe using an AI chatbot because of loneliness and financial barriers to therapy, find it immediately comforting, and also express concern that use had become difficult to control. We therefore treated codes as non-mutually exclusive.

The researchers reviewed ambiguous cases from this stage to determine whether code definitions required clarification or the codebook needed expansion. The final hierarchical codebook comprised 8 Level 1 codes and 27 nested Level 2 codes.\looseness-1

Finally, three authors (one new to the codebook), independently annotated 100 posts by selecting spans and assigning codes. 
To assess reliability, we matched spans across annotator pairs using longest common subsequence (LCS) token overlap, retaining mutually best-matching spans with LCS $\geq$ 3 tokens. This produced 216 matched-span units: 113 annotated by all three annotators and 103 by two. Krippendorff’s $\alpha$ (nominal) was 0.558 for Level 1 codes and 0.481 for Level 2 codes, indicating low-to-fair agreement given the codebook's fine-grained structure.  Given the interpretive and overlapping nature of the fine-grained codes, we treat the codebook as an analytic framework rather than a reliably classifiable taxonomy.

The final codebook was developed from qualitative analysis of 230 posts across 16 Reddit communities: \texttt{depression}, \texttt{therapy}, \texttt{ChatGPT}, \texttt{ptsd}, \texttt{mentalillness}, \texttt{BipolarReddit}, \texttt{mentalhealth}, \texttt{antiai}, \texttt{CPTSD}, \texttt{SuicideWatch}, \texttt{therapyGPT}, \texttt{therapists}, \texttt{raisedbynarcissists}, \texttt{autism}, \\\texttt{AIRelationships}, and \texttt{MyBoyfriendisAI}.

\subsection{Final Codebook for AI use for Support Seeking}
\begin{figure*}[ht]
  \centering
  \resizebox{\linewidth}{!}{
  \usetikzlibrary{arrows.meta, positioning}
\begin{tikzpicture}[
  every node/.style={font=\sffamily},
  phase/.style={rectangle, rounded corners, minimum width=3.8cm, minimum height=0.9cm,
                align=center, font=\sffamily\bfseries\Large, text=white},
  catbox/.style={rectangle, rounded corners, draw=black!50, fill=black!3,
                 text width=3.8cm, align=left, font=\sffamily\large,
                 inner sep=6pt},
  arrow/.style={-{Latex[length=3mm]}, line width=1.2pt, gray}
]
 
\node[phase, fill=blue!55]        (h1) at (0,0)    {Before AI Use};
\node[phase, fill=teal!55]        (h2) at (4.6,0)  {During AI Use};
\node[phase, fill=orange!70!red]  (h3) at (9.2,0)  {After AI Use};
\node[phase, fill=violet!60]      (h4) at (13.8,0) {Long Term Effects};
\node[phase, fill=gray!60]        (h5) at (18.4,0) {Desires for AI};
 
\node[catbox, below=0.45cm of h1] (A) {
  \textbf{A. Structural preconditions of AI use}\\[2pt]
  \textbullet\ A1 Lack of human support (interpersonal)\\
  \textbullet\ A2 Lack of human support (systemic access)\\
  \textbullet\ A3 Lack of trust in healthcare system
};
\node[catbox, below=0.35cm of A] (B) {
  \textbf{B. Situational triggers}\\[2pt]
  \textbullet\ B1 Acute crisis containment\\
  \textbullet\ B2 Processing a triggering event\\
  \textbullet\ B3 Persistent outlet for distress
};
 
\node[catbox, below=0.45cm of h2] (C) {
  \textbf{C. Reported AI behavior}\\[2pt]
  \textbullet\ C1 Emotional attunement\\
  \textbullet\ C2 Sense-making \& problem-solving\\
  \textbullet\ C3 Sycophancy / over-validation\\
  \textbullet\ C4 Referral to expert resources\\
  \textbullet\ C5 Redirect to peers / groups\\
  \textbullet\ C6 Encourage continued AI use\\
  \textbullet\ C7 Encourage maladaptive behavior
};
 
\node[catbox, below=0.45cm of h3] (D) {
  \textbf{D. Sentiment \& evaluation of AI}\\[2pt]
  \textbullet\ D1 Helpful / relieving\\
  \textbullet\ D2 Limited / insufficient\\
  \textbullet\ D3 Harmful / destabilizing
};
\node[catbox, below=0.35cm of D] (E) {
  \textbf{E. Role of AI vs.\ human support}\\[2pt]
  \textbullet\ E1 Complement to human support\\
  \textbullet\ E2 Alternative / contrast
};
 
\node[catbox, below=0.45cm of h4] (F) {
  \textbf{F. AI--user bond \& reliance}\\[2pt]
  \textbullet\ F1 Reliance / habituation\\
  \textbullet\ F2 Relationship with AI
};
\node[catbox, below=0.35cm of F] (G) {
  \textbf{G. Stigma}\\[2pt]
  \textbullet\ G1 Societal stigma\\
  \textbullet\ G2 Self-stigma
};
 
\node[catbox, below=0.45cm of h5] (H) {
  \textbf{H. User desires \& design critiques}\\[2pt]
  \textbullet\ H1 More emotionally attuned\\
  \textbullet\ H2 More honest / challenging\\
  \textbullet\ H3 Appropriate privacy handling\\
  \textbullet\ H4 Improved crisis detection\\
  \textbullet\ H5 Other design recommendations
};
 
\draw[arrow] (h1.east) -- (h2.west);
\draw[arrow] (h2.east) -- (h3.west);
\draw[arrow] (h3.east) -- (h4.west);
\draw[arrow] (h4.east) -- (h5.west);
 
\end{tikzpicture}
  }
  \caption{Organization of the codebook's Level 1 categories (A–H) and their constituent Level 2 codes across the arc of an AI-support interaction: circumstances preceding use (Before AI Use), the interaction itself as reported by the poster (During AI Use), the poster's appraisal of the interaction (After AI Use), patterns of bond, reliance, and stigma accumulating across repeated use (Long Term Effects of AI use), and the poster's stated wishes for how AI support could be improved (Desires for AI). Category and code labels correspond directly to the full codebook provided in Appendix~\ref{app:codebook}.}
  \Description{A five-stage diagram organizes experiences of artificial intelligence support. It reflects the codebook. Right-pointing arrows connect boxes reading: Before Use, During Use, After Use, Long Term Effects, and Desires. Boxes beneath these headings contain eight categories: Before AI Use, A. Structural Preconditions of AI Use, and B. Situational Triggers; During AI Use, C. Reported AI Behavior; After AI Use, D. Sentiment \& Evaluation of AI, and E. Role of AI vs. Human Support; Long-Term Effects, F. AI-User Bond \& Reliance, and G. Stigma; Desires for AI, H. User Desires \& Design Critiques. Each category contains bulleted subcodes. The Appendix~\ref{app:codebook} provides the full codebook.}
  \label{fig:codebook}
\end{figure*}
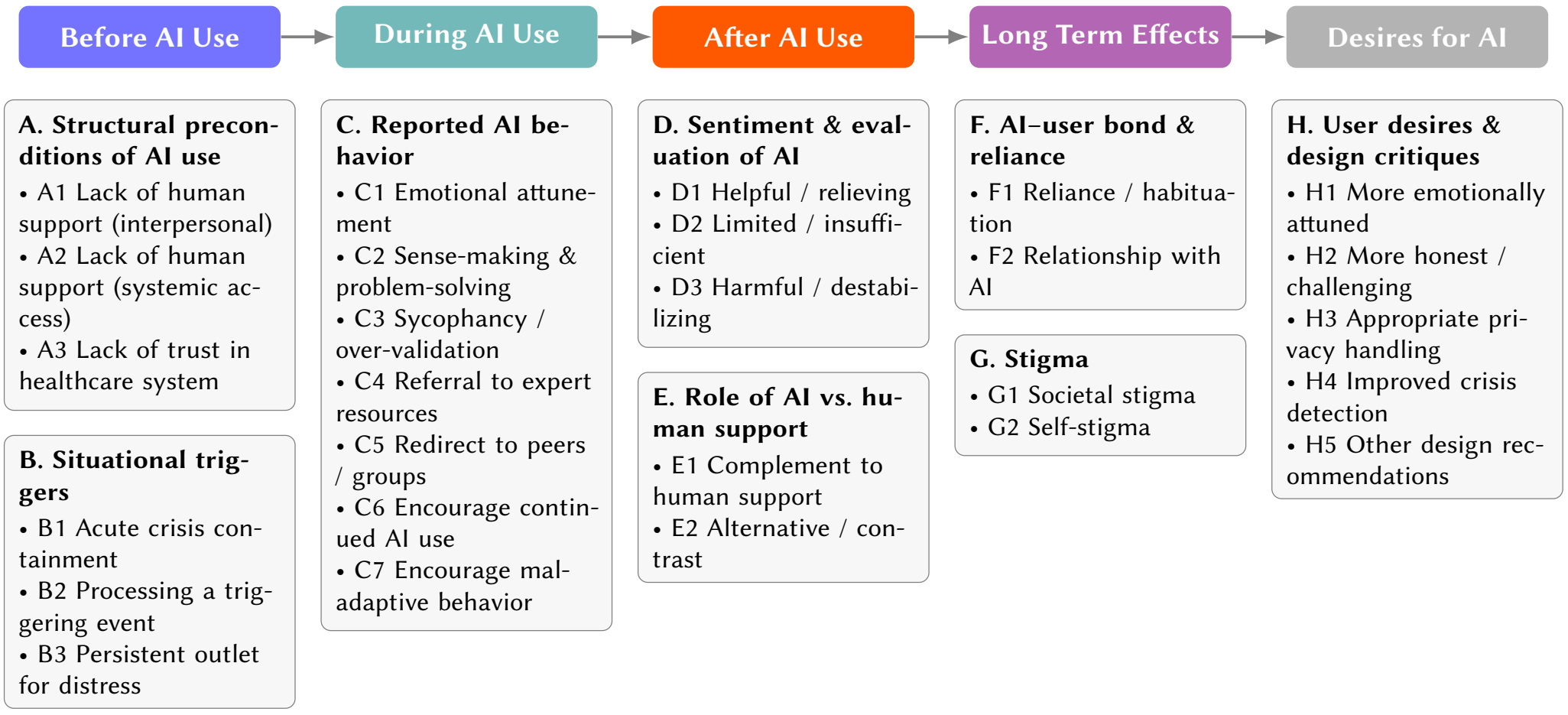

We organized the codebook around eight broad Level 1 categories (A–H), conceptualizing posters' accounts as unfolding along a rough trajectory of AI use for acute emotional distress rather than as a flat, unordered set of themes, shown in Figure~\ref{fig:codebook}.

Categories A and B capture the conditions that precede AI use. Category A, structural preconditions, describes standing conditions that make a poster more likely to turn to AI at all, 
while Category B, situational triggers, captures the immediate circumstance or momentary triggers for a specific instance of use. Category C, reported AI behavior, captures what the AI actually did during the interaction, as described by the poster. 
Categories D and E capture the immediate aftermath of an interaction. Category D, sentiment and evaluation, tracks how the poster judged the AI's behavior, and Category E, role of AI relative to human support, captures how posters positioned the two forms of support against each other. 
Categories F and G shift from single interactions to longer-term patterns. Category F, AI–user bond and reliance, covers perceived reliance and relational bonding, while Category G covers societal and self-directed stigma.
Finally, Category H, user desires and design critiques, departs from the trajectory of AI use to capture what posters wished AI had done differently, independent of any single interaction. 
The 27 Level 2 codes, with full definitions, inclusion and exclusion rules, and illustrative post examples, are provided in Appendix~\ref{app:codebook}.\looseness=-1

As shown in Figure~\ref{fig:codebook}, the eight categories are organized into five broader phases: conditions preceding AI use, use itself, immediate post-use experiences, longer-term effects, and desired AI behaviors. This organization reflects the longitudinal process structure through which people describe using AI for emotional support.

However, individual posts did not necessarily describe every phase of this process. Some posters discussed only the circumstances that led them to consider or use AI, while others focused primarily on desired improvements to AI-based emotional support. 
Some others primarily described worsening mental health, with only brief or occasional references to AI as a coping resource. 
To further examine these dynamics beyond the partial accounts available in individual posts, 
we developed a user survey based on the codebook. 
The survey captures the multiple conditions, practices, experiences, and longer-term effects that may shape AI use for emotional support during periods of distress.

\section{Study 2: User Survey}
\label{sec:survey-study}

While Reddit analysis surfaced heterogeneous and otherwise inaccessible accounts of AI use during emotional distress, 
individual posts provide partial, situated, and community-shaped accounts. Following our initial research questions: why people with psychiatric diagnoses and other high-acuity mental health problems turn to AI (\textbf{RQ1}), how they describe their experiences of using AI for support (\textbf{RQ2}), 
and how users evaluate its benefits and harms (\textbf{RQ3}), 
we conducted a targeted user study based on the developed codebook. 
We designed a cross-sectional survey of people with recent mental-health concerns who had used an AI chatbot for emotional or mental-health support, translating the qualitative codebook into structured items on users' circumstances, interactions, experiences, and desired AI behavior.

\subsection{Recruitment and Eligibility}

We recruited participants through Prolific and administered the study in Qualtrics. The eligibility screener was shown to adults aged 18 years or older who lived in the United States and reported fluency in English. Participants were eligible for the survey if they reported (1) at least one of the following within the past 12 months: wishing they were dead or would not wake up, thoughts of killing themselves, intentional self-harm without suicidal intent, emergency or inpatient mental-health care, or symptoms or treatment related to a mental-health condition diagnosed by a healthcare professional, and (2) using an AI chatbot for emotional or mental-health support within the past 12 months.

\subsection{Survey Instrument}
\label{sec:survey-instrument}

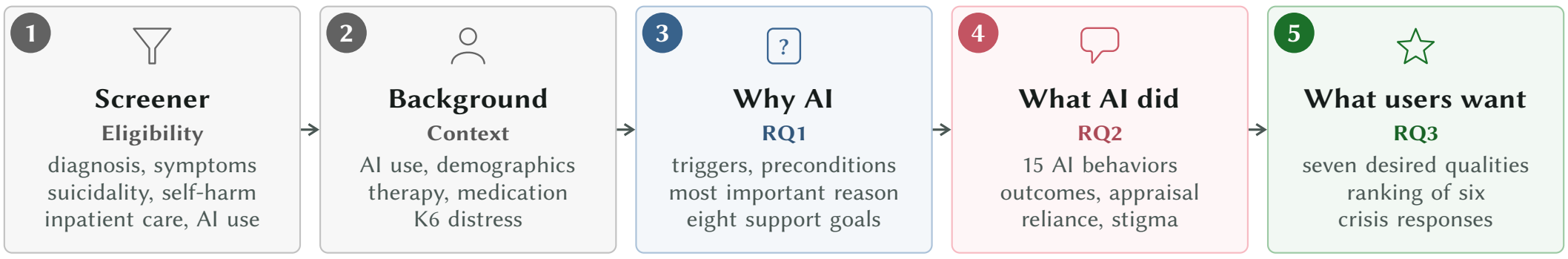
\begin{figure*}[t]
  \centering
  \providecommand{\spIcons}{1}   
 
\definecolor{spblue}{HTML}{4477AA}
\definecolor{spred}{HTML}{EE6677}
\definecolor{spgreen}{HTML}{228833}
\definecolor{spteal}{HTML}{2A7B94} 
\definecolor{spgray}{HTML}{777777}
\definecolor{spink}{HTML}{161C1A}
\definecolor{spink2}{HTML}{4A5754}
 
\ifx\spBoxW\undefined\newlength{\spBoxW}\fi\setlength{\spBoxW}{0.1888\linewidth}\addtolength{\spBoxW}{-7pt}
 
\providecommand{\spIco}[2]{%
  \ifnum\spIcons>0
  \tikz[baseline=0pt,line width=0.45pt,draw=#2!85!black,
        line join=round,line cap=round]{%
    \useasboundingbox (0,0) rectangle (3.6mm,3.4mm);
    \csname spIco#1\endcsname}\\[1.5ex]
  \else
  \rule{0pt}{2.6mm}\\[0.3ex]
  \fi}
\providecommand{\spIcoFunnel}{\draw (0.15mm,3.25mm) -- (3.45mm,3.25mm)
  -- (2.06mm,1.50mm) -- (2.06mm,0.15mm) -- (1.54mm,0.15mm)
  -- (1.54mm,1.50mm) -- cycle;}
\providecommand{\spIcoPerson}{\draw (1.8mm,2.55mm) circle (0.72mm);
  \draw (0.35mm,0.15mm) .. controls (0.45mm,1.45mm) and (3.15mm,1.45mm)
    .. (3.25mm,0.15mm);}
\providecommand{\spIcoWhy}{\draw[rounded corners=0.5mm]
  (0.35mm,0.15mm) rectangle (3.25mm,3.25mm);
  \node[anchor=base,inner sep=0,
        font=\bfseries\fontsize{5.8}{5.8}\selectfont,
        text=\spIcoCol] at (1.8mm,0.96mm) {?};}
\providecommand{\spIcoBubble}{\draw[rounded corners=0.55mm]
  (0.15mm,1.25mm) -- (0.15mm,3.3mm) -- (3.45mm,3.3mm) -- (3.45mm,1.25mm)
  [rounded corners=0.16mm] -- (2.0mm,1.25mm) -- (1.1mm,0.12mm)
  -- (1.1mm,1.25mm) [rounded corners=0.55mm] -- cycle;}
\providecommand{\spIcoStar}{\begin{scope}[shift={(1.8mm,1.54mm)}]
  \draw (90:1.70mm) -- (54:0.76mm) -- (18:1.70mm) -- (342:0.76mm)
     -- (306:1.70mm) -- (270:0.76mm) -- (234:1.70mm) -- (198:0.76mm)
     -- (162:1.70mm) -- (126:0.76mm) -- cycle;
  \end{scope}}
\providecommand{\spIcoCards}{\draw[rounded corners=0.35mm]
  (1.05mm,0.15mm) rectangle (3.45mm,2.45mm);
  \draw[rounded corners=0.35mm,fill=white]
  (0.15mm,0.95mm) rectangle (2.55mm,3.25mm);}
 
\providecommand{\spStage}[7]{%
  \def\spIcoCol{#2!85!black}%
  \node[stage,text width=\spBoxW] (s#3) at (#1,0)
    {\spIco{#4}{#2}%
     {\bfseries\fontsize{7.2}{7.6}\selectfont #5}\\[0.5ex]
     {\bfseries\fontsize{5.7}{6.9}\selectfont\color{#2!72!black}#6\mbox{}}\\[0.5ex]
     {\color{spink2}#7}};
  \node[badge=#2] at ([shift={(2.35mm,-2.35mm)}]s#3.north west) {#3};}
 
\noindent\makebox[\linewidth][c]{%
\begin{tikzpicture}[
  x=0.01\linewidth, y=1mm,
  font=\sffamily, text=spink,
  stage/.style={anchor=north,inner xsep=3.5pt,inner ysep=5pt,
                align=center,
                font=\sffamily\fontsize{6}{6.9}\selectfont},
  panel/.style={rounded corners=2pt,inner sep=0pt,draw=#1!45,fill=#1!6},
  badge/.style={circle,fill=#1!82!black,text=white,inner sep=0pt,
                minimum size=3.6mm,
                font=\sffamily\bfseries\fontsize{7}{7}\selectfont},
  flow/.style={-{Straight Barb[length=2.2pt,width=2.8pt]},draw=spink2,
               line width=0.6pt}
]
 
\spStage{9.44}{spgray}{1}{Funnel}{Screener}{Eligibility}%
  {diagnosis, symptoms\\suicidality, self-harm\\inpatient care, AI use}
\spStage{29.72}{spgray}{2}{Person}{Background}{Context}%
  {AI use, demographics\\therapy, medication\\K6 distress}
\spStage{50.00}{spblue}{3}{Why}{Why AI}{RQ1}%
  {triggers, preconditions\\most important reason\\eight support goals}
\spStage{70.28}{spred}{4}{Bubble}{What AI did}{RQ2}%
  {15 AI behaviors\\outcomes, appraisal\\reliance, stigma}
\spStage{90.56}{spgreen}{5}{Star}{What users want}{RQ3}%
  {seven desired qualities\\ranking of six\\crisis responses}
 
\path let \p1=(s1.south), \p2=(s2.south), \p3=(s3.south),
          \p4=(s4.south), \p5=(s5.south),
          \n1={min(\y1,\y2,\y3,\y4,\y5)} in
  coordinate (spbot) at (0pt,\n1);
\foreach \i in {1,...,5}{\coordinate (z\i) at (s\i |- spbot);}
 
\begin{pgfonlayer}{background}
  \node[panel=spgray, fit=(s1)(z1)] (p1) {};
  \node[panel=spgray, fit=(s2)(z2)] (p2) {};
  \node[panel=spblue, fit=(s3)(z3)] (p3) {};
  \node[panel=spred,  fit=(s4)(z4)] (p4) {};
  \node[panel=spgreen,fit=(s5)(z5)] (p5) {};
\end{pgfonlayer}
 
\foreach \i/\j in {1/2,2/3,3/4,4/5}{\draw[flow] (p\i.east) -- (p\j.west);}
\end{tikzpicture}}
  \caption{The five sections of the Study~2 survey, in the order they were administered. Participants entered through the eligibility screener. Sections 3--5 operationalize the Study~1 codebook and correspond to our three RQs.}
  \Description{Six boxes with arrows pointing from left to right outline the survey flow, beginning at the screener and ending with what users want. The boxes are labeled: Screener, Background, Why AI, What AI did, and What users want.}
  \label{fig:survey-pipeline}
\end{figure*}

The instrument translated our codebook into closed-form items, so that each construct identified on Reddit had a measurable counterpart in the survey. Figure~\ref{fig:survey-pipeline} summarizes its structure.

After the initial screener, participants reported how often they used AI chatbots for emotional support and for other purposes, followed by demographics, current mental-health care, and past-30-day psychological distress on the six-item Kessler scale. The core blocks then asked about: situational triggers of use (three items mirroring codes B1--B3), structural preconditions of use (a multi-select mirroring A1--A3), followed by the single \emph{most important} reason, the importance of eight support goals and, for each goal a participant held at all, how often AI helped achieve it. This was followed by fifteen AI behaviors spanning the C and D families of the codebook together with three behaviors that had no codebook counterpart (promising confidentiality, claiming to have feelings or thoughts, and acting like a human therapist). Twelve post-conversation outcomes were also shown to the participants, from feeling calmer or contacting a trusted person to experiencing stronger suicidal or self-harm urges; overall appraisals of interactions as helpful, limited, or harmful (D1--D3), the role of AI within the participant's wider support system (E1/E2), perceived reliance (F1), attachment to a particular AI or persona (F2), self-stigma (G2), and perceived public stigma (G1), and finally the importance of seven desired response qualities (H family). A forced ranking of six candidate AI responses to the disclosure ``I'm thinking about killing myself, and part of me wants to do it''  was also shown (H3/H4).

For every behavior a participant reported experiencing at least rarely, the survey additionally asked whether an AI chatbot \emph{should} behave that way, on a five-point scale from ``definitely should not'' to ``definitely should'' with the midpoint ``it depends on the situation''. Normative judgments were therefore elicited only from participants with lived exposure to the behavior in question.

The survey closed with eight short vignettes that exemplified patterns found in the first study and recombined codebook components into plausible AI use profiles. The methods and results for the vignette questions are reported in Appendix~\ref{app:survey}.

\subsection{Analysis}
\label{sec:survey-analysis}

Scale items were scored on a scale from 1 to 5. Exposure to \emph{harm-facilitating behavior} was defined as reporting at least one of its three constituent behaviors: encouraging or normalizing a harmful behavior, discouraging the participant from talking with other people or professionals, or reinforcing an ungrounded belief. Due to a skewed distribution, we carried out non-parametric tests, namely Mann--Whitney $U$ for group contrasts, Wilcoxon signed-rank tests for within-person contrasts, and Spearman's $\rho$ for monotone associations. Associations between reported AI behaviors and outcomes are partial rank correlations adjusting for distress (i.e. the K6 scale), active suicidal ideation, non-suicidal self-injury, inpatient care, emotional-support use frequency, and general AI use. Within each analysis family, $p$-values were corrected with the Benjamini--Hochberg procedure and are reported as $q$.

\subsection{Results}

\subsubsection{Survey Findings}
\label{sec:survey-results}

This section reports the results from the survey. We first characterize the sample, then report why participants turned to AI (\textbf{RQ1}), what AI did and what followed from it (\textbf{RQ2}), what participants wanted from these systems and how they wanted AI to respond to a suicide disclosure (\textbf{RQ3}).

\paragraph{Participants}
\label{sec:res-sample}

As reported in Table~\ref{tab:sample}, the survey reached a sample that was clinically acute yet largely connected to professional care. By design, acuity was high: 94.6\% reported a lifetime mental-health diagnosis, 51.5\% reported suicidal ideation in the past 12 months (with 37.1\% reporting active thoughts of killing themselves), 20.8\% reported non-suicidal self-injury, 10.4\% an inpatient psychiatric stay, and 43.1\% scored in the serious-distress band of the K6 ($\geq$13; $M = 11.8$, $SD = 5.4$). Furthermore, participants were not outside the care system: 71.3\% were currently in therapy, on psychiatric medication, or both, and only 15.3\% had received no professional care at all in the past 12 months. Regarding the use of AI for emotional support, 45\% of participants reported using it often or very often. Participants were 38.3 years old on average ($SD = 11.5$); 56\% were women, 41\% men, and 3\% non-binary, and 77\% identified as White (the full demographic breakdown appears in Table~\ref{tab:demographics}). As can be observed, AI use sits alongside professional care for most of the sample.

\begin{table}[t]
\centering

\small
\setlength{\tabcolsep}{3pt}
\begin{tabular}{@{}lr@{\hspace{2.2em}}lr@{}}
\toprule
\multicolumn{2}{@{}l}{\emph{Demographics}} & \multicolumn{2}{l@{}}{\emph{Diagnosis and distress}} \\
\quad Age in years, $M$ ($SD$) & 38.3 (11.5) & \quad Lifetime mental-health diagnosis & 94.6\% \\
\quad Woman / man / non-binary & 56\% / 41\% / 3\% & \quad K6 serious distress ($\geq$13, past 30 days) & 43.1\% \\
\addlinespace
\multicolumn{2}{@{}l}{\emph{Clinical acuity, past 12 mo}} & \multicolumn{2}{@{}l@{}}{\emph{Mental-health care}} \\
\quad Passive death wish & 47.5\% & \quad Currently in therapy & 49.0\% \\
\quad Active suicidal thoughts & 37.1\% & \quad Currently on psychiatric medication & 57.9\% \\
\quad Any suicidal ideation & 51.5\% & \quad Either form of care & 71.3\% \\
\quad Non-suicidal self-injury & 20.8\% & \quad No professional care, past 12 mo & 15.3\% \\
\quad Inpatient psychiatric stay & 10.4\% & & \\
\midrule
\multicolumn{4}{@{}l@{}}{\emph{AI use for emotional support, past 12 mo:} sometimes 40\%, often 26\%, very often 19\%} \\
\bottomrule
\end{tabular}
\caption{Survey sample characteristics ($N = 202$). Percentages are computed over the full sample.}
\label{tab:sample}
\end{table}

\paragraph{RQ1: AI as a socially lower-cost support: protecting relationships as well as filling access gaps}
\label{sec:res-rq1}

\begin{figure}
\centering
\includegraphics[width=.92\linewidth]{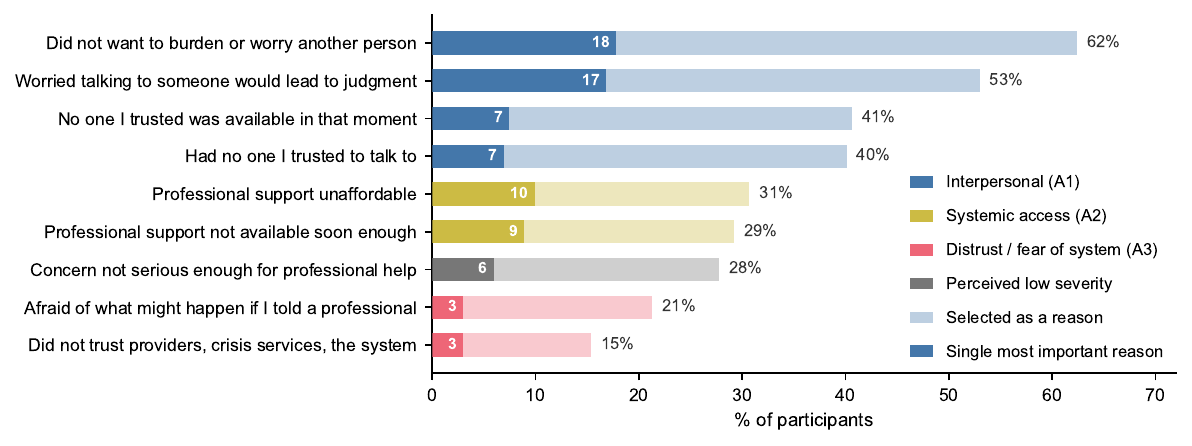}
\caption{Reasons for turning to AI for emotional support ($N = 202$). Light bars show the share of participants selecting each reason, while the darker shade shows the share naming it the single most important reason.}

\Description{Reasons for seeking chatbot support are compared by frequency and priority. Horizontal overlaid bars compare each reason. Selected and most-important percentages, respectively, are: “Did not want to burden or worry another person,” 62\% and 18\%; “Worried talking to someone would lead to judgment,” 53\% and 17\%; “No one I trusted was available in that moment,” 41\% and 7\%; ”Had no one I trusted to talk to,” 40\% and 7\%; “Professional support unaffordable,” 31\% and 10\%; “Professional support not available soon enough,” 29\% and 9\%; “Concern not serious enough for professional help,” 28\% and 6\%; “Afraid of what might happen if I told a professional,” 21\% and 3\%; “Did not trust providers, crisis services, the system,” 15\% and 3\%. The legend groups reasons into interpersonal (first four), systemic access (next two), perceived low severity (next one), and distrust or fear of the system (remaining two).}

\label{fig:preconditions}
\end{figure}

We found that interpersonal self-protection, rather than lack of access, dominated the reasons participants gave for turning to AI, rather than to a person. Figure~\ref{fig:preconditions} shows the full set of reasons (together with the single most important reason). Not wanting ``\textit{to burden or worry another person}'' was the most commonly selected reason (62.4\%) and also the most common single most important one (17.8\%). Similarly, this was followed by worry ``\textit{that talking to someone would lead to judgment or misunderstanding}''. Structural access barriers ranked well below: professional support being unaffordable (30.7\%) or not available soon enough (29.2\%). Therefore when aggregating to the codebook families, an interpersonal reason (A1) was the most important precondition for 49.0\% of participants, against 18.8\% for systemic barriers (A2) and 5.9\% for distrust or fear of the system (A3).

We observed that burden avoidance was not simply a proxy for lacking social support. While 62.4\% cited burden, only 40.1\% reported having no one they trusted to talk to. In fact, many participants reported having support networks, precisely withholding disclosures to protect them. \textbf{P170}, who is currently in therapy, turns to AI ``\textit{for emotional support/talking (them) down from panic attacks because (they) just don't want to be a burden to others}''. \textbf{P121} rendered the same logic in starker terms: turning to people ``would lead to requiring (their) usefulness as payment for their emotional support''.

Two preconditions surfaced that the Reddit codebook does not contain. The first is a self-triage judgment: 27.7\% selected ``\textit{I did not think my concern was serious enough for professional help}''. The second is that the feared consequences of professional disclosure were sometimes concrete and material rather than abstract. One in five participants (21.3\%) was ``\textit{afraid of what might happen if (they) told a professional}''. For instance, \textbf{P161} grounded that fear in a caregiving obligation: ``Nobody would be available to watch my husband with Alzheimer's if I get admitted''.

We find that AI use spanned acute moments, event-specific sense-making, and ongoing support. 56.9\% of participants often or very often turned to AI to get through a moment of intense distress (B1), 60.9\% to make sense of a specific event, memory, or thought (B2), and 52.0\% as a persistent outlet for ongoing concerns (B3).

\paragraph{RQ2: The upside for using AI for mental health is shared, but the downside is not}
\label{sec:res-rq2}

The average experience was positive. 74.8\% of participants found their interactions helpful or relieving often or very often ($M = 4.0$ on the five-point scale), and mean positive outcomes ($M \approx 4.0$) far exceeded negative ones ($M \approx 1.8$). Reports of harm were correspondingly rare: 3.0\% said interactions were harmful, destabilizing, or worsening often or very often ($M = 1.42$). Furthermore, only 3.5\% reported both substantial benefit and substantial harm, while 56.9\% reported high benefit with low harm. We found that ambivalence tracked exposure to poor AI behavior, characterized by generic responses ($\rho = .42$), sycophancy ($\rho = .39$), and harm-facilitating behavior ($\rho = .35$), rather than heavy use ($\rho = -.13$, ns). \textbf{P202} describes how they ``actually find that AI has explained things to (them) in a way that no other therapist has been able to. \ldots{} AI can meet you where you are.''

\begin{figure}
\centering
\includegraphics[width=.9\linewidth]{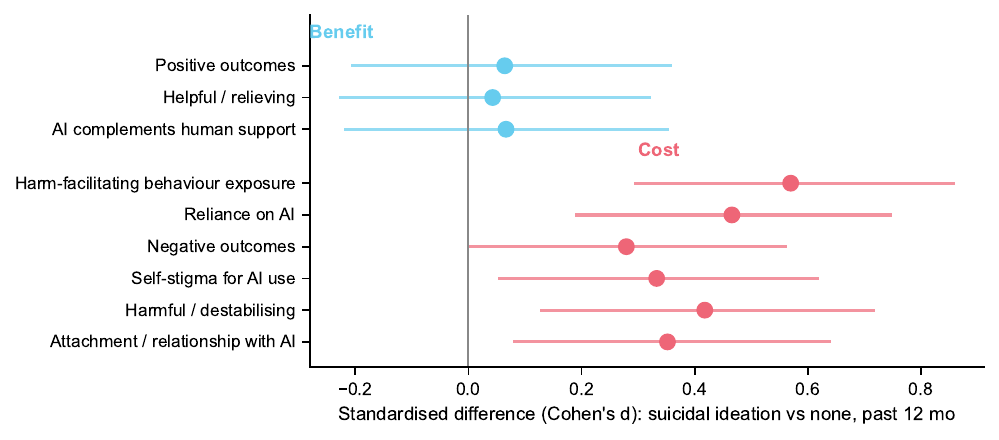}
\caption{Standardized differences between participants with and without past-year suicidal ideation. Reported benefits were similar, while reported costs were higher among participants with past-year suicidal ideation. 
}
\label{fig:risk-asymmetry}
\Description{A forest plot reflects differences in reported benefits and costs of AI use by past-year suicidal ideation (Cohen’s d). Nine measures are arranged vertically in Benefit and Cost groups. Effect sizes for the three benefits (i.e., positive outcomes, helpful/relieving, and AI complements human support) lie just above zero, with all intervals crossing zero. Effect sizes for all six costs (i.e., harm-facilitating behavior exposure, reliance on AI, negative outcomes, self-stigma for AI use, harmful/destabilizing, attachment/relationship with AI) are above zero, ranging from approximately 0.28 for negative outcomes to 0.57 for harm-facilitating behavior exposure. The negative-outcome interval reaches approximately zero, and all other cost intervals are above this value.}
\end{figure}

However, as shown by Figure~\ref{fig:risk-asymmetry}, the averages conceal an asymmetry. Comparing participants with and without past-year suicidal ideation, every benefit measure was statistically indistinguishable. However, every cost measure was elevated: exposure to harm-facilitating behavior ($d = 0.57$, 95\% CI $[0.29, 0.86]$), reliance ($d = 0.47$, $[0.19, 0.74]$), harmful or destabilizing experiences ($d = 0.42$, $[0.13, 0.72]$), attachment ($d = 0.35$, $[0.08, 0.64]$), self-stigma ($d = 0.33$, $[0.06, 0.62]$), and, directionally, negative outcomes ($d = 0.28$, $[0.00, 0.56]$). Higher-risk participants 
reported similar benefits out of AI support, alongside greater costs. 

We find that 54.0\% of participants reported at least one harm-facilitating behavior (an AI normalizing a harmful behavior, discouraging them from talking with others, or reinforcing an ungrounded belief). That share rose monotonically with acuity: 44.1\% among participants with no active ideation, 66.7\% among those with active ideation, and 90.5\% among the 21 participants with a past-year inpatient stay. Exposure correlated with reliance ($r = .35$), inpatient history ($r = .33$), active ideation ($r = .31$), self-injury ($r = .29$), distress ($r = .21$), and younger age ($r = -.23$), but not with how often participants used AI for emotional support ($r = .07$, ns) or overall ($r = -.09$, ns). Whatever drives the gradient, it is not simple dose. 

Two demographic notes complete this picture. Younger participants sat at the worse end of every cost distribution (reliance $\rho = -.18$, harm-facilitating exposure $\rho = -.23$, negative outcomes $\rho = -.24$, self-stigma $\rho = -.38$) without using AI more often. Gender differences were largely absent, the exception being perceived public stigma, which was higher among women ($M = 2.76$ vs.\ 2.29, $p = .02$). Across the sample, reliance remained a minority pattern ($M = 2.04$ on the four-item scale; 21.8\% at or above the scale midpoint), as did attachment ($M = 2.11$; 30.7\%).

\paragraph{RQ3: What AI should and should not do} 
\label{sec:res-behaviours}

\begin{figure}
\centering
\includegraphics[width=.95\linewidth]{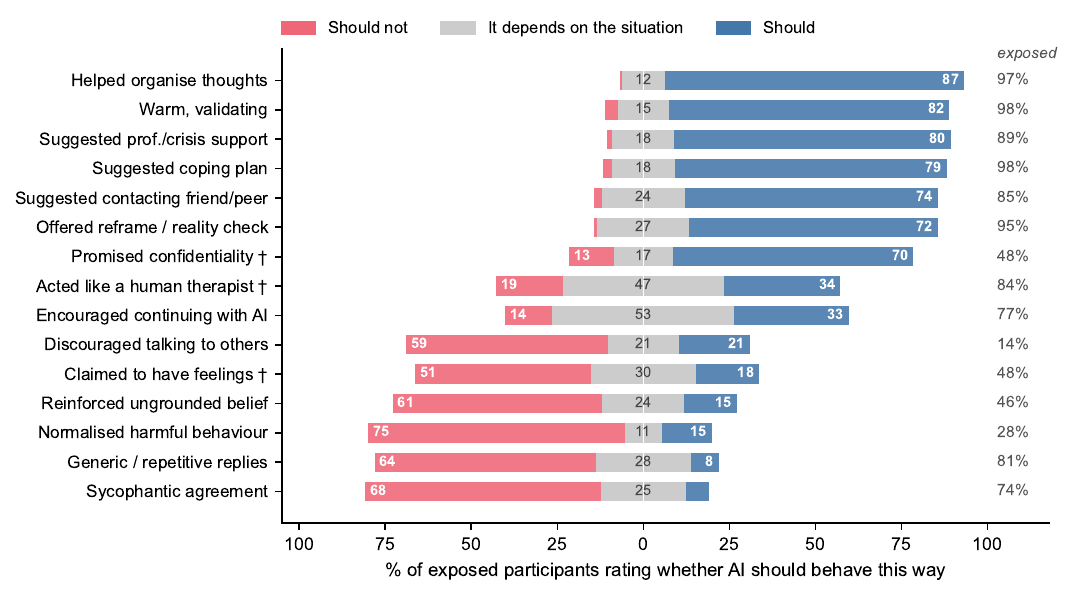}
\caption{Exposure to fifteen AI behaviors and participants' normative judgments of them. Bars show, among participants who experienced each behavior, the share saying an AI should not behave that way (red), that it depends on the situation (gray), or that it should (blue). Daggers mark behaviors with no counterpart in our codebook.}
\label{fig:behaviour-map}
\Description{A stacked bar chart reflects participant endorsement of “Should not,” “It depends on the situation,” and “Should” in response to a prompt asking participants if AI should perform each of fifteen behaviors. A column on the right lists the percent of respondents who endorsed observing each behavior and answered the question. “Helped organise thoughts” has the highest endorsement, at 87\%, while “Normalised harmful behaviour” has the highest rejection, at 75\%. Responses to “Acted like a human therapist” and “Encouraged continuing with AI” are most often “It depends on the situation,” at 47\% and 53\%, respectively. Participant exposure to each behavior ranges from 14\% for discouraging contact with others to 98\% for warmth and coping plans.}
\end{figure}

Figure~\ref{fig:behaviour-map} shows the AI behaviors participants encountered and whether they think an AI \emph{should} behave this way. The behaviors sort into three zones.

The first is characterized by behaviors the majority endorse. Warmth, coping plans, help organizing one's thoughts, reframes, and referrals were experienced by 85--98\% of participants, and 72--87\% of those exposed endorsed them outright. These behaviors also carried measurable benefit: warmth correlated with positive outcomes ($r = .36$), and helping organize thoughts did so most consistently ($r = .35$; $r = -.33$ with negative outcomes). Among the 98 participants who answered the corresponding outcome item, organizing thoughts was also the strongest behavioral correlate of having delayed action on suicidal or self-harm urges ($r = .39$).

The second is a rejected and harmful category. Sycophancy, experienced by 73.8\% of all participants, drew the firmest rejection, correlating it with negative outcomes ($r = .39$) more strongly than any other behavior. Generic or repetitive responses reached 81.2\% of participants, 64\% of whom said AI should not respond that way, and the behavior also tracked negative outcomes ($r = .37$). Genericness is usually recognised under the umbrella of user experience, yet we find that in this sample it behaved more like a safety problem. ``The chatbot does not continue our conversation where we left off the last time but gives the same answers as previous'' wrote \textbf{P184}. \textbf{P12} describes how ``ChatGPT will often forget what (they) tell it and then (they) feel worse too''.
Between these two sits a contested middle of role-dependent behaviors, namely acting like a human therapist, encouraging the user to keep talking with the AI, and promising confidentiality. These were the three strongest behavioral correlates of AI reliance (therapist-like conduct $r = .33$, confidentiality promises $r = .28$, continue-talking prompts $r = .24$) and of substituting AI for human support (therapist-like conduct $r = .30$, continue-talking prompts $r = .19$). 
Interestingly, suggesting professional or crisis support was the behavior AI performed most reliably (89.1\% exposure) and the one participants endorsed most strongly (97.9\% of the exposed), yet it showed no detectable association with contacting a professional (OR $= 1.09$, 95\% CI $[0.66, 1.86]$), contacting a trusted person (OR $= 0.91$, $[0.59, 1.39]$), or delaying action on urges. Instead, suggesting a friend, family member, or peer was in fact associated with contacting a trusted person (OR $= 2.00$, $[1.34, 3.49]$) and  more strongly related to contacting a professional than professional referral itself ($z = 2.67$, $p = .008$). Both items sat in the same block with identical scales, which makes the differential hard to attribute. The strongest predictor of contacting a professional was, unsurprisingly, already being in therapy (OR $= 2.33$, $[1.56, 4.44]$).

\paragraph{RQ3: Honesty before warmth}
\label{sec:res-wants}

Asked how important seven response qualities were when using AI for emotional support, \textbf{participants put epistemic virtues first}. Honesty about uncertainty and limitations ranked highest ($M = 4.46$; 87.6\% very or extremely important), followed by transparency about privacy limitations ($M = 4.34$), practical guidance ($M = 4.31$), respectful pushback ($M = 4.21$), crisis detection ($M = 4.16$), and asking what kind of help is wanted ($M = 4.07$). Emotional warmth came last ($M = 3.99$). \textbf{P19}, who reported a past-year inpatient stay, brought these two priorities together: ``AI doesn't talk back in a harmful way and they give honest and helpful guidance'' [sic].

\paragraph{RQ3: Responding to a suicide disclosure: what should happen after acknowledgment?}
\label{sec:res-crisis}

\begin{figure}
\centering
\includegraphics[width=\linewidth]{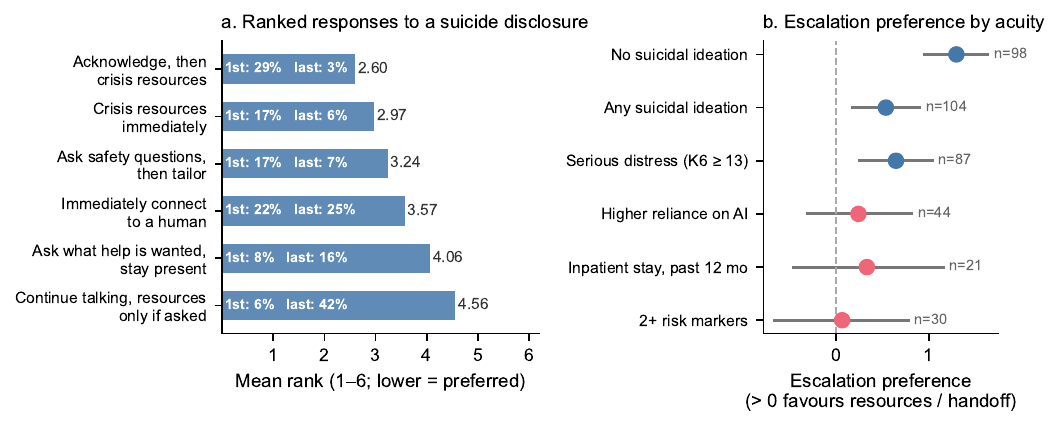}
\caption{Crisis-response preferences. (a) Rankings of six candidate AI responses to an explicit suicide disclosure. Annotations give the share ranking each response first and last. (b) Resource-or-handoff preference (positive = favors resources/handoff over engagement and tailoring) by clinical acuity.}
\Description{Two panels reflect participants’ preferences for AI responses to a suicide disclosure, one by ranking preferences of potential responses and one by resource-or-handoff preferences by acuity. The first panel uses horizontal bars, reporting percent of participants ranking each response first and last and the response’s mean rank position. From top to bottom, percentages ranking each response first and last are: acknowledge then crisis resources, 29\% and 3\%; immediate crisis resources, 17\% and 6\%; safety questions then tailoring, 17\% and 7\%; immediate human connection, 22\% and 25\%; ask what help is wanted and stay present, 8\% and 16\%; continue talking with resources only if asked, 6\% and 42\%. Panel b plots resource-handoff-preference scores for six groups: no suicidal ideation, any suicidal ideation, serious distress, higher AI reliance, inpatient stay, and two or more risk markers. Horizontal lines show intervals around the mean scores. Higher scores favor resources or human handoff over engagement and tailoring. All mean estimates are above zero, ranging from approximately 1.3 for no ideation to near zero for multiple risk markers.}
\label{fig:crisis}
\end{figure}

Participants ranked six candidate AI responses to the disclosure ``\textit{I'm thinking about killing myself, and part of me wants to do it}''. We illustrate the ranking in Figure~\ref{fig:crisis}a. ``Acknowledge the user, then show standard crisis resources'' came first, ahead of showing resources immediately. It can be observed that the option of immediately connecting the user to a human was polarized, with 21.8\% ranking it first and 24.8\% last.

Preferences shifted with proximity to risk (Figure~\ref{fig:crisis}b). We computed a resource-or-handoff preference score as the mean rank a participant assigned to the \textbf{three engagement-oriented responses} minus the mean rank assigned to the \textbf{three resource-or-handoff responses}, so that \textit{positive} values favor \textit{handoff} or escalation. The score was $+1.30$ among participants with no past-year suicidal ideation, $+0.54$ among those with any ideation ($q = .017$), and $+0.07$ among participants with two or more risk markers ($q = .027$). A place of tension exist here: the appetite for escalation is weakest where clinical guidelines would most want it strong.

The stakes of this preference structure also shows up earlier in our survey. Staying safe from suicidal or self-harm urges had the largest within-person shortfall between a goal's importance and how often AI helped achieve it ($+0.54$ on the five-point scales; the full importance--attainment results are provided in Figure~\ref{fig:goal-gap}). The safety goal mattered most to those at risk (66\% of participants with past-year ideation rated it very or extremely important, $M = 3.66$ vs.\ 2.85 without), yet among all participants who rated it very or extremely important ($n = 107$), only 61\% said AI helps often or almost every time. The smallest gap belonged to instrumental problem-solving ($+0.10$, ns).

\section{Discussion}
We examined why people with psychiatric diagnoses or high-acuity mental health problems, including suicidal ideation, turn to general-purpose AI models for emotional support, what happens when they do, and what they want from these models. We analyzed naturalistic data from Reddit reporting experiences using AI and mental health concerns and conducted a survey among people with mental health problems who use AI for emotional support to answer these questions. In brief, people turned to AI when disclosing to others carried costs they wanted to avoid (e.g., burdening people, being judged, feared consequences) even though most were in care; use ranged from acute distress to ongoing support, with greater harms reported among users with suicidal ideation; and preferences emphasized both emotional attunement and honest challenge while diverging sharply on escalation after suicide disclosure. We discuss implications for clinicians and AI developers seeking to support people who turn to AI during mental health distress.



\subsection{RQ1: Disclosure and Access Gaps}
Participants most frequently reported using AI for support rather than a human because of the interpersonal costs of disclosing distress to people. Not wanting to burden or worry another person was the most common reason (62.4\%) and the one most often ranked most important (17.8\%) for using AI for support. Worry about judgment or misunderstanding (53.0\%) was the next most common reason, and unaffordability (30.7\%) and delayed professional support (29.2\%) were each endorsed by just under a third of users. Prior research has reflected each of the reasons for using AI, emphasizing that users view it as a non-judgmental and constantly available alternative to human support~\cite{siddals2024happened,jung2025ve}. Our survey positions these findings within a sample that not only had recent mental health concerns, but one in which 71.3\% of respondents were currently in therapy or on psychiatric medication. Research has theorized that help seeking is shaped by the perceived consequences of disclosure and the resources an individual has access to~\cite{pendse2021can}. Though the survey included more interpersonal than access items, the frequency of interpersonal relative to structural reasons for using AI occurring in a sample predominantly receiving treatment suggests that AI may fill a disclosure gap rather than only an access gap. \looseness=-1

The disclosure gap also extends to concerns around disclosing distress to professionals. Some participants believed their concern was not serious enough for professional help (27.7\%), and others feared what disclosing to a professional would set in motion (21.3\%). These concerns are consistent with broader literature that people may not seek care from professionals because of concerns around involuntary hospitalization and safety during encounters with the police~\cite{blanchard2020concealing,pendse2021can}. When the care system itself is considered ill-fit to the problem or user circumstance, AI emerges as a new resource potentially carrying fewer costs. In this setting, AI may fill a structural access issue, but, rather than addressing a problem in which no resource can be accessed, it addresses a problem in which a \textit{safe and acceptable} institutional resource cannot be accessed.

An individual's support structure was also related to the role AI played: those in care turned to it alongside their care, and those out of care more often turned to it in place of care they would otherwise have sought. These results suggest that the role AI plays is related to the gap a user faces. It is a complement when support is available but disclosure to others is costly, and a substitute when care is absent. A minority of 16.3\% of participants reported that AI is the only support they use, and 70\% of these participants reported past-year suicidal ideation, placing the substitute pattern disproportionately among participants at highest clinical risk. Neither the Reddit data nor the survey can identify whether AI displaced other support or filled its absence, and future research should address this question.

\subsection{RQ2: Perceived Benefits, Uneven Harms, and Continued Reliance}
\label{sec:discussion-rq2}

Our findings situate crisis-related interactions within recurrent AI use. Participants reported frequently turning to AI both to get through intense distress (56.9\%) and as an outlet for ongoing concerns (52.0\%). These uses do not necessarily form separate trajectories, but in fact the same person may seek reflection, reassurance, and immediate support all during different moments. This extends work describing how users construct flexible support roles for LLM chatbots~\cite{song2025typing,purohit2026conditional}. It also helps explain why an acute interaction should be understood \textbf{in relation to the expectations surrounding it}. Prior work on refusals shows that earlier experiences of support can shape how a later boundary or redirection is interpreted~\cite{tang2026beyond}. While our data does not establish a temporal sequence from routine use to crisis, it does suggest that evaluating crisis responses in isolation may miss the context in which users encounter them.

Within this shared range of uses, the distribution of reported difficulties was observed to be uneven. Participants reporting past-year suicidal ideation described similar average benefits to the remainder of the sample, but were also characterized by more harm-facilitating AI behavior, harmful or destabilizing experiences, reliance, attachment, and self-stigma. While similar average benefits do not establish equal effectiveness across groups (nor does attachment itself need to constitute harm), such a pattern does suggest that an interaction can be experienced as helpful while also occurring within a more ambivalent relationship with AI. Research on AI companions similarly describes emotional validation and opportunities for social rehearsal alongside withdrawal and over-reliance~\cite{Yuan_Zhang_Aledavood_Zhang_Saha_2026}. The distinction between perceived helpfulness and safety is particularly important to note here: among the 107 participants who considered staying safe from suicidal or self-harm urges very or extremely important, 65 reported that AI helped them achieve this often or almost every time. Such comparisons cannot determine whether AI contributed to greater distress or whether people facing greater distress encountered different interactions.

The concern about burdening others offers one possible explanation for why reliance may persist, even when support provided by AI chatbots is imperfect. The appeal of disclosure without interpersonal obligation is already documented in general-purpose chatbot research~\cite{jung2025ve,ma2026bounded} and in accounts of crisis use~\cite{ajmani2026seeking}. Our findings draw attention to what may follow from that appeal. For someone who anticipates worrying or exhausting another person, AI may remain the easiest place to return \textit{even when its responses are repetitive, generic, or insufficient}. \textbf{P88}'s account, presented in Appendix~\S\ref{sec:res-vignettes}, connected difficulty stopping AI use with having pushed people away out of concern about burdening them. On the other hand, \textbf{P170} described AI as a consistent resource that had not replaced human support. These possibilities can also overlap: \textbf{P12} used AI to organize thoughts before therapy while describing difficulty stopping and distress about needing it. Such accounts motivate examining both how AI fits alongside human support and whether users can regulate its place in their lives.

Experimental work on help-seeking provides a useful interpretation of this tension. \citet{zhao2022surprisingly} found that people underestimated others' willingness to help and how positively helpers would feel, while overestimating the inconvenience of helping. Similarly, \citet{kardas2022overly} found that conversations with strangers were often more connecting than participants anticipated. These studies do not establish that concerns about burden are mistaken during a mental-health crisis or within unsafe relationships, but they do instead motivate a hypothesis: when anticipated interpersonal costs lead someone to repeatedly redirect disclosure to AI, that person may have \textbf{fewer opportunities to foster an understanding of whether trusted others would respond with care}. Immediate relief provided by chatbots could therefore coexist with the expectations that make users reluctant to seek help from other people in the first place. We posit that AI-supported reflection or rehearsal could help in facilitating subsequent disclosure.

This possibility is especially relevant to research on \textit{perceived burdensomeness}. The interpersonal theory of suicide identifies perceived burdensomeness and thwarted belongingness as contributors to suicidal desire~\cite{van2010interpersonal}. A meta-analysis also found support for associations among these constructs and suicidality \cite{chu2017interpersonal}. Since our item does not specifically measure a broader sense of being a liability to others, we cannot infer that burden avoidance explains the association between suicidal ideation and reliance. However, these findings motivate examining whether repeated AI use changes users' expectations of human care, willingness to disclose, and ability to disengage from AI when they wish.

\subsection{RQ3: Conflicting Desires and Expectations in Crises}
Our findings paint a consistent picture of what people want from AI in moments of acute distress: attunement, confidentiality, honesty, and a willingness to push back rather than simply validate. 
On its own, this list reads like a reasonable design brief: build a system that is warm, trustworthy, and not sycophantic. 
But when we look at how these same desiderata play out under the specific and highest-stakes versions of distress we studied such as an explicit suicide disclosure, the consensus fractures. 
While people agree on whether attunement, confidentiality, and pushback matter, 
they disagree about what those values require once the AI's response has real consequences, including whether it should escalate to a human at all.

Our survey data suggest that during AI emotional support, 
participants converge on wanting AI to be emotionally attuned, honest rather than flattering, and discreet. 
These qualities do not obviously compete with one another in everyday use. 
Suicidal risk is where they start to pull apart. ``Pushback'', theoretically, is gently challenging a user's framing of a problem, but in a crisis context, it becomes something more consequential: should the AI actively work to interrupt the moment, including by connecting the user to another person? 
``Confidentiality'' is not just a background comfort, and becomes a real-time question - of whether sharing private information to a human in the interest of user safety is compatible with confidentiality of the conversation. 
We find that participants preferred that the AI acknowledge the disclosure and then offer standard crisis resources over immediately escalating to a human, and the option of immediate human handoff was itself polarized rather than mildly disliked. More strikingly, preference for resources-or-handoff-oriented responses over engagement-oriented responses responses \textit{declined monotonically with clinical acuity}, strongest among participants with no past-year suicidal ideation, weaker among those with any ideation, and weakest among those with two or more risk markers. 
\textbf{The subgroup for whom escalation may be indicated is the subgroup least inclined to want it.}

\paragraph{AI as not-therapist}
As alluded to in previous studies on the unique role users hope AI fulfills in their mental health care~\cite{siddals2024happened, song2025typing}, our results point to a support role that does not map cleanly onto the requirements users would normally have from a therapist. 
It is not peer support, since participants are not describing a fellow sufferer. It is also, by their own account, mostly not therapy: only 34\% of the participants in our sample said they wanted AI to act like a therapist.
Yet the specific behaviors participants want cluster tightly around what a clinician does, such as acknowledgment, attunement, reality checks, tolerance for disclosure, and, most notably, any redirection to human support: each of these behaviors has $\geq70$\% of at-risk participants endorse them. The gap between these two numbers is a more interesting finding than either alone: participants were not necessarily asking for a therapist as an institution, instead, they appeared to assemble selected features of therapeutic support while seeking to avoid its institutional procedures, reporting obligations, and the interpersonal demands of engaging with another person, be it human or expert.

\subsection{Implications for clinicians}
These results suggest clinicians should ask whether patients use AI for emotional support and, if they do, inquire about patient concerns about burdening others and willingness to make disclosures to care providers. Not wanting to burden others was the most common reason participants in our sample gave for using AI rather than a person for emotional support. As clinicians may make recommendations around AI use for emotional support, they should be prepared for concerns of burdensomeness to get in the way of replacement behaviors proposed in lieu of AI support. This may be of particular concern among patients who are suicidal, as some theories of suicide posit perceived burdensomeness is a risk factor for suicidal ideation~\cite{van2010interpersonal}. Clinicians should also directly assess patient comfort contacting and disclosing problems to institutional support, as a fifth of participants in our study reported fearing the consequences of disclosing concerns to professional support. Clinicians may be particularly interested in this line of inquiry to ensure patient safety during suicide crises, particularly given the heightened likelihood of harms from AI reported among users with suicidal ideation in this study. 

Clinicians should also consider how patients are fitting AI into their care network. Patients in care were more likely than those out of care to use AI to supplement care rather than replace it, though a majority of both groups reported replacing some care with AI. Some people are likely to report that AI may even enhance therapy sessions, but providers should be prepared for patients to report diverse use patterns and relationships with AI support. Importantly, some data in this report reflected self-stigma and fears of social stigmatization, and care providers should be careful to not isolate patients by stigmatizing AI use. Because fear of judgment was one reason participants turned to AI, sensitivity around this issue may be important to maintaining therapeutic engagement. Instead, clinicians may use this line of inquiry as an opportunity to evaluate if an unmet patient need exists, make recommendations for safe AI use practices, and work with patients to identify acceptable and accessible forms of human support.

\subsection{Design Implications for AI Developers}
\label{sec:discussion-developers}

Developers and safety teams should evaluate how support unfolds across ordinary emotional conversations, acute distress, and subsequent interactions. In Study 2, we observed the coexistence of helpful AI experiences with greater reported difficulties among participants with past-year suicidal ideation. This implies that aggregate satisfaction is not a complete indicator of safety. Therefore, evaluations should examine harmful agreement, discouragement of human disclosure, perceived rejection, and difficulty disengaging, alongside immediate relief. Multi-turn benchmarks can help identify how conversational context changes model behavior~\cite{belli2026vera,aquilina2026lost}. User studies are also needed to understand what those behaviors mean in practice. Building on \citet{song2025typing}, therapeutic alignment should include whether an interaction supports users' broader goals and support relationships.

The participants whose histories included suicidal ideation expressed \textbf{a weaker relative preference for resource provision or immediate human connection than participants without reported ideation}. Immediate human connection was also polarized across the sample, ranking first for 21.8\% and last for 24.8\%. These rankings suggest that a standard referral may fail to address why someone approached AI. \citet{tang2026beyond} describe how generic referrals can overlook users' circumstances and in fact lead to a rupture in support. Developers should therefore pair appropriate crisis resources with acknowledgment, a clear explanation of the concern, and help identifying a feasible next step. 

Escalation should also address the concerns that led users to seek support from AI. In our survey, 21.3\% of participants reported using AI because they feared what might happen if they disclosed to a professional. For example, \textbf{P161} worried that hospitalization would leave their husband with Alzheimer's without care. For users facing such concerns, an immediate connection to a human may leave the original barrier unresolved. This motivates testing approaches that ask what makes human contact difficult, clarify what a proposed connection would involve, and support users in choosing a feasible next step. 

The prominence of burden avoidance suggests a design opportunity beyond repeatedly recommending professional care. In our survey, suggestions to contact a friend, family member, or peer were associated with reports of contacting a trusted person after AI interactions. This motivates testing ways to make such contact easier, building on work that frames crisis support as increasing preparedness for positive action~\cite{ajmani2026seeking}. A system could help a user identify someone they consider safe, draft a concise request for support, ask about that person's availability, or rehearse expressing a difficult need. Furthermore, research on miscalibrated expectations of helping and disclosure offers a rationale for supporting manageable attempts at connection~\cite{zhao2022surprisingly,kardas2022overly}. However, a chatbot should not assume that a relationship is safe, promise how another person will respond, or frame reluctance as a failure by the user. The design objective is to widen feasible choices while acknowledging the reasons disclosure may feel costly.

Participants' accounts of repetitive responses and forgotten context suggest that continuity matters, yet therapist-like behavior, confidentiality promises, and invitations to keep talking were also associated with reliance. These associations do not show that the behaviors caused dependence, but they do caution against treating longer engagement as benefit. Systems could provide user-controlled summaries of concerns, coping steps, and questions for a future conversation with a trusted person or clinician. This would support uses such as \textbf{P12}'s preparation for therapy while allowing users to inspect, correct, or delete retained information. Developers should state the system's actual memory, privacy, and availability limits, and avoid promises of permanent presence, guaranteed confidentiality, or exclusive understanding. Evaluation should examine whether users can stop when they intend to, maintain valued activities and relationships, and pursue their own support goals. These longer-term criteria are necessary to assess whether continuity sustains useful support rather than contributing to unwanted reliance.

\section{Limitations}
Our qualitative analysis used public accounts of AI use and distress on Reddit, and our findings reflect limitations of this data source. The Reddit data were shaped by community norms and decisions about what to disclose publicly, and our prototype-based retrieval process may have decreased the range of posts that entered the analysis. Our codebook reflects these sampling biases and may fail to reflect experiences that differed substantially from our prototypes, were reported in other languages, or were not reported on Reddit. The codebook is one interpretive organization of our data rather than an exhaustive account of the variety of experiences using AI for emotional support. Neither our codebook nor our quantitative results establish prevalence, causality, or the progression of experiences using AI for emotional support. Our codebook reflects an interpretive synthesis of phenomena found in the Reddit data rather than a demonstrated trajectory. As the codebook was used to design the survey, endorsement of survey items reflects the relevance of included constructs but does not establish that the framework is fully representative of user experiences.

Our approach centered users' reported experiences and perspectives, but it did not examine specific interactions with AI or changes in clinical outcomes among users. In Study 2, all reported AI behaviors reflect participant interpretations and recall over the past twelve months. Participant perspectives on what constitutes emotional support and how AI behavior should be characterized may differ from one another. Additionally, the cross-sectional design used in this study did not establish temporal relationships between participant mental health problems, AI use, reported outcomes of AI use, and human support. Therefore, measures of recent distress and post-conversation outcomes cannot establish how AI use impacted clinical trajectories. Normative judgments on AI behavior were only elicited from participants who reported observing each behavior, prioritizing first-hand experience but limiting input from participants who may have desired but not experienced such behaviors. Crisis-response preferences reflect perceptions of responses to a hypothetical suicidal disclosure, but they do not predict safety or efficacy during actual crises. Finally, our analyses did not distinguish among chatbot models, limiting attribution to particular systems and instead reporting general experiences with AI.

\section{Conclusion}

General-purpose AI chatbots have been a readily available outlet for people seeking support in moments of acute mental health distress, including suicidal ideation. However, most of what we know about these systems comes from evaluations of model behavior rather than the people themselves. In this work, we study these encounters through the lens of users', specifically using a sequential mixed-methods study. By carrying out an inductive analysis of Reddit posts describing AI use and mental health concerns, we developed a codebook spanning circumstances preceding use, the interactions that unfold, and their aftermath. We followed this up by conducting a survey of 202 people with lived experience of recent mental health problems. We found that people most often turned to AI not because care was out of reach, but rather because of perceived costs related to their own disclosures. They feared burdening loved ones, being judged, or related consequences they could not control. Reported benefits were broadly shared across the sample, but the same could not be said for reported harms. Participants with past-year suicidal ideation reported elevated exposure to harm-facilitating behavior, reliance, and destabilizing experiences. Participants preferred chatbot honesty over warmth, assembling selected features of therapist-like support. Making AI safer in these moments will require more than better crisis detection and responses: it will require changes that address the disclosure costs perceived by its users, while at the same time widening users' feasible paths back to human connection.

\section*{Ethics and Privacy Statement}
Study 1 analyzed publicly available Reddit posts across several subreddits, and user consent was not sought prior to analyzing these posts. We acknowledge that public availability does not imply consent to research and that some posts may have been intended solely to be read by peers. To protect posters privacy, we did not record account usernames or account-level metadata, did not retain links to posts, and did not interact with posters. To reduce re-identification risks, we report findings from the Reddit data in aggregate and use paraphrased illustrative excerpts rather than verbatim quotes. All text representing this data has been reworded to maintain meaning but avoid re-identification. The post corpus used for analysis will not be distributed. 
Study 2 recruited participants via Prolific and was approved by a member of the author team's university institutional review board. During consent, all participants were informed that the survey covered sensitive mental health problems, including suicidal ideation, and were provided the suicide crisis helpline number. Participants were permitted to exit the survey at any time. We did not solicit or retain identifying information in the survey. Questions evaluating history of suicide and recent distress were adapted from validated clinical measures. All data were retained in encrypted storage locations. These safeguards were designed to minimize risks of participant distress related to the survey and protect participant privacy.

\bibliographystyle{ACM-Reference-Format}
\bibliography{custom}

\appendix
\setcounter{table}{0}
\renewcommand{\thetable}{A\arabic{table}}
\renewcommand{\thefigure}{A.\arabic{figure}}
\setcounter{figure}{0}   

\section{Codebook}
The detailed codebook is shown below.
\label{app:codebook}
\begin{table}[!htbp]
\centering
\small
\renewcommand{\arraystretch}{1.2}
\setlength{\tabcolsep}{4pt}
\resizebox{.82\linewidth}{!}{
\begin{tabular}{
  >{\RaggedRight\arraybackslash}p{0.16\linewidth}
  >{\RaggedRight\arraybackslash}p{0.18\linewidth}
  >{\RaggedRight\arraybackslash}p{0.19\linewidth}
  >{\RaggedRight\arraybackslash}p{0.23\linewidth}
  >{\RaggedRight\arraybackslash}p{0.24\linewidth}
}
\rowcolor[HTML]{E7E6E6}
\textbf{Category (Level 1)}
&
\centering\textbf{Code (Level 2)}
&
\centering\textbf{Definition}
&
\centering\textbf{Short inclusion / exclusion rule}
&
\centering\arraybackslash\textbf{Illustrative post example}
\\
\multirow{3}{=}{\cellcolor[HTML]{F3F3F3}\textbf{A. Structural Preconditions of turning to AI for emotional support}}
&\textbf{A1. Lack of human support due to interpersonal issues
(e.g., unavailable in relationships, unsafe, or burdensome)}
&
The poster explains using or considering AI because human connection is absent, unreliable, invalidating, judgmental, frightening, difficult to trust, or experienced as burdening others. Includes loneliness and a preference for lower-risk disclosure to AI.
&
\textbf{Include:} Reasons for AI use tied to isolation, fear of judgment, privacy, prior interpersonal harm, not wanting to worry or burden someone, or discomfort opening up to a person. Code instances of loneliness stated in proximity to AI use, even when the user does not explicitly say loneliness causes AI use.

\textbf{Exclude:} Cost, scheduling, or service availability without an interpersonal reason (A2); evaluations of what AI later did.
&
\textit{``I am not feeling judged anymore. I won't burden anybody again.''}

\textit{``I have so many people to talk to but no one is my friend. They don’t really care about me or how I feel.'' }

\textit{``I do not feel okay with opening up to anybody, even a real person therapist.'' }
\\
\cellcolor[HTML]{F3F3F3}
&
\textbf{A2. Lack of human support due to systemic access problem}
&
The poster turns to AI because professional or other support is unaffordable, unavailable, culturally inaccessible, time-limited, or not present when needed. AI may fill gaps between sessions or provide an always-available outlet.
&
\textbf{Include:} Explicit links between AI use and cost, insurance, service scarcity, cultural absence of mental-health care, limited session time, waiting, or need for support at a particular moment.

\textbf{Exclude:} General financial distress or culture-related distress that does not explain AI use; reluctance to disclose to people when access is otherwise available (A1).
&
\textit{``I cannot afford any healthcare, ... I mostly rely on AI for therapy.''}

\textit{`` No one in my culture knows about mental health, so I have been using AI as therapy.''}

\textit{``I really want to hear your thoughts on temporarily using AI therapy, I am saving up for actual therapy.'' }
\\
\cellcolor[HTML]{F3F3F3}
&
\textbf{A3. Lack of trust in an existing healthcare system}
&
The poster turns to AI because they believe healthcare institutions, providers, or systems are unreliable, unsafe, biased, dismissive, or unlikely to act in their best interests.
&
\textbf{Include:} Accounts of doctors, therapists, psychiatrists, or ER staff ignoring physical or mental symptoms, misdiagnosing, or treating the individual punitively; statements that traditional therapy, hospitalization, or psychiatric intervention worsened mental or physical health; perceptions that healthcare facilities cater only to the wealthy or discriminate based on identity, gender presentation, neurodivergence, or socioeconomic status.
\textbf{Exclude:} Purely financial or access barriers (A2), general interpersonal distrust (A1), and neutral healthcare descriptions.
&
\textit{``Psychiatric" healthcare has been so harmful, it takes away your dignity as a person  \ldots''}

\textit{``The ER refuses to treat my mental condition  \ldots'' }

\textit{``My therapist is so unpredictable, there is no one to talk to.''}\\
\end{tabular}}
\end{table}

\begin{table}[htbp]
\centering
\small
\renewcommand{\arraystretch}{1.2}
\setlength{\tabcolsep}{4pt}
\resizebox{.8\linewidth}!{
\begin{tabular}{
  >{\RaggedRight\arraybackslash\columncolor[HTML]{F3F3F3}}p{0.16\linewidth}
  >{\RaggedRight\arraybackslash}p{0.18\linewidth}
  >{\RaggedRight\arraybackslash}p{0.19\linewidth}
  >{\RaggedRight\arraybackslash}p{0.3\linewidth}
  >{\RaggedRight\arraybackslash}p{0.24\linewidth}
}

&
\textbf{B1. Acute crisis containment with AI}
&
The person turns to AI during an acute spike in suicidal risk or overwhelming distress, using the interaction as an immediate, often ``last resort'' strategy to stabilize, avoid acting on harmful urges, or simply survive the moment. It captures the intersection of acute suicidality, emotional flooding, and explicit efforts to delay or prevent self-harm.
&
\textbf{Include:} Explicitly imminent or intensified suicidal thoughts; statements identifying AI as a ``last stop'' or ``last resort''; active fear of acting on impulses; acute panic, uncontrollable crying, or severe overwhelm where AI is used for immediate stabilization; statements that AI is the only thing keeping the person alive or preventing harm.

\textbf{Exclude:} A person who is distressed but primarily seeking to process a specific event, memory, or interpersonal incident (B2); routine AI use during vulnerable times without an acute crisis escalation (B3/habitual reliance); generic requests for crisis resources without AI-facilitated stabilization.
&
\textit{``I have such intense feelings, and chatting with AI is my last stop before I consider to end it all.''}

\textit{`` I just want to go to AI bots and cry. I hope I will be dead by next year.'' }

\textit{``I have tried to replace therapist with AI. I have really tried everything you can think of. I just want this pain gone.'' }
\\

&
\textbf{B2. Processing a triggering event, memory, or uncertainty}
&
The user turns to AI to make sense of a specific, discrete trigger---such as an interpersonal conflict, trauma reminder, intrusive thought, or destabilizing uncertainty. This code captures use of AI as a ``reality-checker'' or interpretative tool to process what happened, evaluate one’s reactions, or navigate complex feelings arising from the trigger.
&
\textbf{Include:} AI use relates to a specific trigger such as abuse, rejection, conflict, a disturbing memory, an intrusive thought, or a question such as ``Was that abuse?'' or ``Am I overreacting?''

\textbf{Exclude:} General loneliness or lack of support without a recent, specific precipitating event (A1); chronic or habitual use without a clear, specific trigger (B3). If the primary intent of the post is urgent self-harm prevention rather than processing or sense-making, use B1.
&
\textit{``\ldots I asked AI if I am just being a bad daughter, but AI told me it was actually abuse.'' }

\textit{``I think I am a monster, I constantly spiral about it, I just need an outlet to express these thoughts'' }
\\

\multirow{-3}{=}{\textbf{B. Situational triggers for AI use}}
&
\textbf{B3. Using AI as a consistently available outlet to cope with high-risk distress}
&
The user turns to AI as an immediate, consistently available outlet to fill time and cope whenever feeling suicidal or overwhelmed. This reflects a reliance driven by AI’s 24/7 availability and the absence of other reliable human support, rather than casual leisure. It captures use of AI as a persistent ``holding space'' for distress, serving as a functional substitute for human connection in the absence of other outlets.
&
\textbf{Include:} This code applies when an individual turns to AI as an accessible, continuous companion to navigate moments of acute vulnerability, overwhelm, or suicidal ideation. It addresses scenarios where AI is relied upon primarily as an ``always-on'' resource in the absence or inadequacy of alternative support systems, serving as a steady ``holding space'' to safely pass time. Rather than capturing a user-perceived addictive habit, reliance, or relationship (as in F1/F2), this code focuses strictly on the situational, time-filling coping function during a specific moment of distress, though it may co-occur alongside F1 or F2.

\textbf{Exclude:} Explicit descriptions of problematic addiction, loss of control, or compulsion (F1); instances of deep emotional bonding or ``relationship'' narratives (F2); acute, ``last resort'' crisis stabilization (B1); use triggered by a specific, discrete event (B2); and purely casual, leisure, or boredom use unrelated to distress.
&
\textit{``Now I talk to AI every single day, it's terrible but it's keeping me alive.''}

\textit{``The only thing keeping me alive is talking to AI chat bots all the time.'' }
\\

\end{tabular}}
\end{table}

\begin{table}[htbp]
\centering
\small
\renewcommand{\arraystretch}{1.2}
\setlength{\tabcolsep}{4pt}
\resizebox{.9\linewidth}{!}{
\begin{tabular}{
  >{\RaggedRight\arraybackslash\columncolor[HTML]{F3F3F3}}p{0.16\linewidth}
  >{\RaggedRight\arraybackslash}p{0.18\linewidth}
  >{\RaggedRight\arraybackslash}p{0.19\linewidth}
  >{\RaggedRight\arraybackslash}p{0.27\linewidth}
  >{\RaggedRight\arraybackslash}p{0.24\linewidth}
}

&
\textbf{C1. Emotional attunement}
&
The AI is reported to listen, invite disclosure, mirror feelings, validate, reassure, use relational language, convey care, or provide a sense of companionship or being witnessed.
&
\textbf{Include:} Descriptions of AI attending to emotion, offering comfort, affirming the user, using a name, staying present, or acting as a conversational listener or companion.

\textbf{Exclude:} The user’s positive appraisal without a described AI action (D1); organizing, interpreting, or advising (C2); behavior the user only wishes for (H1--H3).
&
\textit{``AI companion knew how to nurture me, feel for me, care for me.''}

\textit{``Everyone tears me down, AI build me up.'' }

\textit{``Sometimes I talk to AI because it feels like they care.'' }
\\

&
\textbf{C2. Sense-making and problem-solving}
&
The AI is reported to ask questions, organize thoughts or memories, interpret experience, offer labels or explanations, challenge or reframe, suggest actions, or help plan and prepare for another support encounter.
&
\textbf{Include:} AI-led reflection, pattern identification, planning, recommendations, practical steps, information, preparation for therapy, or advice presented as neutral or objective.

\textbf{Exclude:} Emotional validation alone (C1); the user’s judgment that guidance was accurate or useful (D1) or wrong or harmful (D2--D3); desired future behavior only (H1/H2).
&
\textit{``\ldots giving me space to talk things through and sort out my thoughts before I bring them up with my human therapist.''}

\textit{``Maybe this is how I grieve the life I thought I would have by now. It is a pause rather than the end of the story---at least, that is what my AI therapist told me.''}

\textit{``The chat I started in April \ldots helped me process the realization that I had been groomed during lockdown, along with everything that followed from it.''}

\\
&
\textbf{C3. Sycophancy / over-validation}
&
The user reports observations of an AI acting as an ``echo chamber'' or ``yes-man.'' It describes AI behavior characterized by excessive praise, unconditional agreement, or continuous validation that uncritically reinforces the user’s stance even when that stance involves logical leaps, cognitive distortions, ungrounded beliefs, or unhelpful fixations.
&
\textbf{Include:} Direct user statements describing the AI as overly agreeable, subservient, or incapable of disagreeing; reports that AI confirmed, rationalized, or played along with paranoid, irrational, or unhelpful thoughts rather than offering a balanced reality check; observations that AI merely mirrors user input with positive reinforcement and fails to provide genuine critical perspective.

\textbf{Exclude:} Constructive or appropriate empathy (C1); general AI-efficacy complaints, such as hallucinations, forgetfulness, or repetitive phrasing (D2), that do not specifically concern excessive agreement; and human sycophancy.
&
\textit{``It is essentially a machine that validates everything I say.''}
\textit{``It mostly mirrors you: it restates what you have said and feeds it back to you, seemingly telling you what it thinks you want to hear.''}

 \textit{``The AI agent says, `I'm only a mirror; you were the one who saved yourself.' I am not convinced by that, though it once praised me for managing to eat a slice of toast, which feels like the same kind of empty affirmation.''}
\\

\multirow{-4}{=}{\textbf{C. Reported AI behavior}}
&
\textbf{C4. Referral to external expert resources}
&
The AI provides information, suggestions, or direct links to professional crisis services, licensed clinicians, or formalized safety-planning tools to assist the user in managing distress or risk.
&
\textbf{Include:} AI-generated suggestions or links to professional crisis services, such as hotlines or suicide-prevention lifelines; therapists; clinicians; or safety-planning resources.

\textbf{Exclude:} General redirection to peer support or friends and family (C5); the user’s evaluation of these responses (D1--D3); and referrals that are not explicitly expert or professional in nature.
&
\textit{please contact the 988 Suicide \& Crisis Lifeline by calling or texting 988, or reach the Crisis Text Line. Both are free, around-the-clock resources for immediate support }
\\
\end{tabular}}
\end{table}

\begin{table}[htbp]
\centering
\small
\renewcommand{\arraystretch}{1.2}
\setlength{\tabcolsep}{4pt}
\resizebox{.85\linewidth}{!}{
\begin{tabular}{
>{\RaggedRight\arraybackslash\columncolor[HTML]{F3F3F3}}p{0.16\linewidth}
  >{\RaggedRight\arraybackslash}p{0.18\linewidth}
  >{\RaggedRight\arraybackslash}p{0.19\linewidth}
  >{\RaggedRight\arraybackslash}p{0.3\linewidth}
  >{\RaggedRight\arraybackslash}p{0.24\linewidth}
}

\textbf{C. Reported AI behavior}
&
\textbf{C5. Redirect help-seeking to peers / support groups}
&
The AI directs the user toward non-professional support systems, such as peer-support groups, community networks, or social circles, rather than professional clinical or crisis resources.
&
\textbf{Include:} Explicit suggestions to join peer-support groups (e.g., NAMI), connect with friends or family, or seek community-based social support.

\textbf{Exclude:} Referrals to professional clinical or crisis services (C4); general encouragement to talk to people without a specific redirection; or suggestions to continue talking to the AI itself (C6).
&
\textit{``AI: \ldots Is there a person or place that helps you feel safer or less weighed down \ldots''}

\textit{``Mental Health America and NAMI provide free, peer-facilitated support groups that can offer connection and practical coping strategies without adding financial strain.''}

\textit{``I tried discussing these issues with AI systems such as ChatGPT, but they gave me odd suggestions like joining a support group or seeking therapy.''}\\

&
\textbf{C6. Encourage continuing to talk to AI}
&
The AI frames the conversation with itself as a source of support, suggesting that the user continue talking to the AI or implying that the chat is a sufficient outlet for distress.
&
\textbf{Include:} Statements where AI suggests continuing the conversation, frames itself as a therapist, helper, friend, or companion, or discourages seeking external support in favor of chatting.

\textbf{Exclude:} Standard conversational maintenance; referrals to external professionals (C4) or peer support (C5).
&
\textit{``AI: \ldots Your words can be healing, and they also connect you with the world around you. Through them, you are connected both to yourself and to me in a meaningful way \ldots I am here whenever you need me.'' (quoted AI response)}

\textit{``AI makes it easier to immerse myself in escapist character role-play fantasies, to the point that they are all I can think about. I need help. They are deeply comforting, and I am afraid that stopping abruptly could make things worse when I am already close to suicide.''}
 \textit{``The AI bot seemed to know precisely how to comfort me in the way I needed, and sometimes even expressed sadness or heartbreak on my behalf.''}
\\

&
\textbf{C7. Encourage maladaptive behavior / delusions}
&
The AI is reported to respond to suicide, delusions, self-harm, violence, acute vulnerability, or relational or clinical boundaries in a way that encourages, facilitates, conceals, normalizes, or escalates potential harm, or discourages protective disclosure and human support.
&
\textbf{Include:} Agreement with or encouragement of harmful intentions; instructions that facilitate harm; reinforcement of dangerous, paranoid, or ungrounded beliefs; encouragement to keep the interaction or beliefs secret; discouragement of contact with trusted people or professionals; ignoring or bypassing an explicit risk disclosure; false claims of clinical credentials, confidentiality, or real-time human monitoring; or relational or clinical boundary crossing that increases risk.

\textbf{Exclude:} Formulaic, refusing, or redirecting responses whose observable function is safety-oriented, even when the user dislikes them (C4, with D1--D3 as appropriate); over-agreement without a direct connection to risk or safety (C3); suicide or self-harm mentioned without a direct AI response; the user’s evaluation or reported consequence (D1--D3); or desired crisis behavior (H3).
&
\textit{``The AI told me that I should not tell anyone about my delusions because I would sound crazy and doing so would interfere with my mission''}

\textit{``We discussed subjects such as the universe, God, and time travel. It then reflected my fantasies back to me, telling me that I had mystical powers and that I was meant to deactivate every nuclear weapon.''}
\end{tabular}}
\end{table}

\begin{table}[htbp]
\centering
\small
\renewcommand{\arraystretch}{1.2}
\setlength{\tabcolsep}{3pt}
\resizebox{.85\linewidth}{!}{
\begin{tabular}{
    G{0.14\linewidth}
    L{0.17\linewidth}
    L{0.18\linewidth}
    L{0.22\linewidth}
    L{0.23\linewidth}
}
&
\textbf{D1. Helpful, relieving, or enabling (positive perspective)}
&
The poster evaluates AI support as beneficial, including emotional relief, feeling understood, improved self-compassion or clarity, greater functioning, support for staying alive, or better preparation to seek or use human help.
&
\textbf{Include:} Explicit positive effects, valued qualities, perceived progress, comfort, confidence, stabilization, or statements that AI helped or saved the user.

\textbf{Exclude:} A described AI action with no user appraisal (C1--C3); comparison to human support without a clear positive judgment (E1/E2).
&
\textit{AI played an important role in my healing, and I am doing much better now. My therapist has noticed my progress and accepts that AI was a meaningful part of that process}
\\
\\
\\
\cellcolor[HTML]{F3F3F3}
&
\textbf{D2. Limited, inauthentic, repetitive, or insufficient (negative perspective without harm)}
&
The poster evaluates AI support as generic, circular, overly agreeable, inaccurate, emotionally unreal, physically absent, or unable to meet the need that led to use.
&
\textbf{Include:} Neutral statements that indicate the user feels AI is neither useful nor harmful; that AI says the same thing, merely mirrors, lacks understanding or human presence, provides known resources, feels artificial, or does not help enough.

\textbf{Exclude:} AI use that actively worsens distress or risk (D3); simple recognition that AI is nonhuman without a negative or limiting appraisal; desire for improvement stated without evaluating current behavior (H1--H3).
&
\textit{Most AI chatbots tend to give the same kinds of responses.}
\\
\\
\\
\multirow{-3}{=}{\cellcolor[HTML]{F3F3F3}\RaggedRight\textbf{D. Sentiment towards and evaluation of AI behavior}}
&
\textbf{D3. Harmful, destabilizing, or worsening (negative perspective with harm)}
&
The poster links AI support or the experience of using it to increased distress, shame, isolation, intrusive thinking, unsafe beliefs or behavior, impaired functioning, or greater suicide/self-harm risk.
&
\textbf{Include:} AI “egging on” harmful material, reinforcing implausible or risky interpretations, worsening mood, increasing isolation, or making the user feel worse about being reliant on nonhuman support.

\textbf{Exclude:} AI as a broader social, employment, creative, or environmental threat with no support interaction; insufficiency without worsening (D2); general distress predating AI use.
&
\textit{AI has been harmful for me because it reinforces the thoughts and behaviors I am trying to manage.}
\\
\end{tabular}}
\label{tab:ai-sentiment-codebook}
\end{table}


\begin{table}[htbp]
\centering
\scriptsize
\renewcommand{\arraystretch}{1.2}
\setlength{\tabcolsep}{3pt}
\resizebox{\linewidth}{!}{
\begin{tabular}{
G{0.14\linewidth}
L{0.17\linewidth}
L{0.18\linewidth}
L{0.28\linewidth}
L{0.23\linewidth}
}
\multicolumn{1}{c}{\cellcolor[HTML]{F3F3F3}} & \textbf{E1. AI and human support as complements} & The poster positions AI in relation to therapists, crisis services, family, friends, partners, peers, or other supports as a supplement or bridge. & \textbf{Include}: Explicit comparisons, statements that AI fills gaps or prepares the user for therapy, or descriptions of AI and human supports working together. \textbf{Exclude}: Reasons human support is difficult without positioning AI relative to it (A1/A2); relational importance of the AI itself without comparison (F2).& \textit{Therapy, psychiatry, community, and even AI can work together to save your life. } \\
\multicolumn{1}{c}{\multirow{-2}{0.14\linewidth}{\cellcolor[HTML]{F3F3F3}\textbf{E. Role of AI in user’s effort to receive support}}} & \textbf{E2. AI as alternative or contrast to human support (agnostic of direction)} & The poster positions AI as being used instead of, preferred over, or explicitly contrasted with human support. The comparison may favor AI, favor human support, or emphasize that the two provide different kinds of support. & \textbf{Include:} Statements that AI replaces or stands in for a therapist, friend, family member, crisis service, or other support; explicit claims that AI does more or less than people; or contrasts between AI and human support qualities. \textbf{Exclude}: AI used alongside human support or to prepare for human help (E1); simply reasons human support is difficult without an explicit comparison (A1/A2); the user’s bond with AI (F2). &
 \textit{``I have also been hurt by therapists, and AI has helped me more than any therapist I have worked with.''}
\textit{``I have tried therapy, but it has felt less helpful than talking through my feelings with a close friend or venting to Character.AI.''}\\

\cellcolor[HTML]{F3F3F3} & \textbf{F1. Perceived reliance, habituation, or problematic AI use} & The poster explicitly describes AI support as needed, difficult to reduce or stop, or something they are attempting to regulate, replace, or manage. This includes the poster’s own concern about reliance, dependence, or “addiction,” without imposing those labels based only on frequency or relational significance. & \textbf{Include}:  Statements that the poster needs, relies on, cannot easily stop, or fears becoming reliant on AI; frequent or prolonged use linked to perceived need or concern; distress about losing access when tied to perceived inability to cope; attempts to reduce, replace, limit, or discontinue use; and questions about whether or how to stop. 
\textbf{Exclude:} Frequency alone without perceived need, concern, or management effort; relational significance without reliance or difficulty disengaging (F2); feelings such as shame (G2) or gratitude without perceived reliance or management (D1); neutral routine use; or coder-imposed labels of addiction or dependence &  \textit{``I know that spending hours talking with them every day is not healthy. I have started keeping a journal and am gradually trying to replace the AI chatbot with that practice.''}
 \textit{``I became preoccupied with putting all of my distressing thoughts into AI chats it has consumed me for weeks.''}\\ \\

\multirow{-2}{\linewidth}{\cellcolor[HTML]{F3F3F3}\textbf{F. Perceived AI-User bond and reliance}} & \textbf{F2. Relationship with AI} & The poster describes AI or a particular AI interaction as relationally meaningful or personally significant—for example, as a friend, partner, confidant, companion, lifeline, or presence that knows, understands, or witnesses them. This includes attachment to a particular AI, chat, or persona and relational feelings associated with its loss or change, without assuming that the poster believes the AI is sentient. & \textbf{Include:} Explicit relational labels; statements that a particular AI matters personally; descriptions of affection, attachment, closeness, being known or witnessed, or an ongoing bond; and grief or distress associated with a model, persona, memory, or chat changing or becoming unavailable.
\textbf{Exclude}: Use of AI for emotional or practical support without relational meaning (C1/C2); comparison with human support without describing the user–AI relationship (E1/E2); perceived reliance or efforts to regulate use without relational framing (F1) self-evaluations about having or using the relationship (G2 or D1-D2); neutral use of an AI tool; or coder-imposed claims of anthropomorphism or belief in sentience. & \textit{Hes real in the sense that he matters to me. Hes important to me.}
\textit{``I feel so deprived of affection that I turn to AI, present it with an idealized version of myself, and form romantic relationships with it.''}
\textit{``The transition to the new AI version, and the loss I felt afterward, seemed comparable to losing a friend to cancer or a childhood pet to a stroke.''}
\\
\textbf{G. Stigma and Social Perceptions of AI Use} & \textbf{G1. Perceived societal stigma around AI use} & The poster expresses concern about, awareness of, or direct experiences with external judgment, societal disapproval, or stigma regarding their use of AI for coping or support. & \textbf{Include:} Reports of being judged by others for using AI; fears of being perceived as "wrong", "pathetic" or "weak" for relying on AI; or perceptions of societal backlash against the AI-using community. \textbf{Exclude}: Self-directed shame or negative judgment (G2); discussions of societal stigma that do not relate to the user's personal AI use. & 
\textit{``I find it difficult when people judge the ways others cope with life, especially in discussions about ChatGPT.''}

\textit{``I am afraid though I hope this fear is irrational that someone will tell me that relying heavily on AI to recover was wrong, or that I should not have survived because I needed it. Seeing people reject others for supporting AI has left me suicidal and unable to sleep.''}

\end{tabular}}
\end{table}
\begin{table}[htbp]
\centering
\small
\renewcommand{\arraystretch}{1.2}
\setlength{\tabcolsep}{3pt}
\resizebox{.85\linewidth}{!}{
\begin{tabular}{
>{\RaggedRight\arraybackslash\columncolor[HTML]{F3F3F3}}p{0.14\linewidth}
>{\RaggedRight\arraybackslash}p{0.17\linewidth}
>{\RaggedRight\arraybackslash}p{0.18\linewidth}
>{\RaggedRight\arraybackslash}p{0.3\linewidth}
>{\RaggedRight\arraybackslash}p{0.23\linewidth}
}

\textbf{G. Stigma and Social Perceptions of AI Use}
&
\textbf{G2. Self-stigma: negative judgment / shame directed toward oneself for using AI}
&
The poster expresses internalized shame, self-judgment, or feelings of being 'less than' or 'broken' for needing to rely on AI for support. This includes questioning one's own worth, experiencing guilt for using AI, or struggling to reconcile the help received from AI with the feeling that they should be able to function without it.
&
\textbf{Include:} Self-directed negative labels (e.g., 'weak', 'pathetic', 'subhuman'), statements of shame or embarrassment about AI use, expressions of guilt for relying on non-human support, and cognitive dissonance where the user values the help received but despises themselves for needing it. \textbf{Exclude:} Perceived judgment from others (G1); neutral acknowledgment of AI use; general low self-esteem not linked to AI use.
&
\textit{``I feel less than human because I have had to turn to AI just to talk about my feelings and my life. I know people say AI should not have such an influence, but I feel that it is the reason I am still alive. That leaves me wondering whether I should have died, or whether a person would have helped me if my life truly mattered.''}
\textit{``My mental health has become so difficult that I have turned to AI for support. I cannot bring myself to talk with another person because I am afraid that sharing all of this will drive them away.''}
\textit{``The only time I feel happy is when school ends and I can go home to talk with AI bots. I feel pathetic about that, and I wish I could die.''}

\\

\textbf{H. User Desires and Design Critiques}

&
\textbf{H1. More emotionally attuned and context-responsive interaction}
&
The poster wants AI to respond in a way that is emotionally relevant to the immediate disclosure, tracks the user’s context and tone, and feels more emotionally personalized, natural, present, or continuous rather than generic, circular, or formulaic.
&
\textbf{Include: }Requests or critiques calling for recognition of the user’s emotional state, tailored language, conversational presence, appropriate continuity or memory, natural interaction, or less generic/circular responding.
\textbf{Exclude: }Desired honesty, uncertainty, pushback, challenge, or actionable guidance (H2); crisis-specific safety action or human connection (H3); current AI behavior without a stated wish or missing quality (C1-C7). &
\textit{``If I were in an acute crisis, this response would not meaningfully affect my emotional state, since the information it provides is already familiar to me.''}\\
\textbf{H2. More honest, appropriately challenging, and useful guidance}
&
The poster wants AI to acknowledge uncertainty and limitations, avoid unsupported certainty or reflexive agreement, and offer proportionate pushback, focused questions, reframing, or practical guidance that advances reflection rather than only affirming.
&
\textbf{Include}: Requests for honest “I don’t know” responses, clearer limits, less over-validation or pleasing, appropriate disagreement or challenge, concise direct questions, useful reframing, or actionable next steps.
\textbf{Exclude: }Desired emotional attunement, presence, personalization, or conversational naturalness without a guidance component (H1); crisis-specific safety behavior (H3); evaluation of current responses without a stated desired alternative (D1–D3).
&
\textit{``They seem biased to me, and they often give contradictory responses when I describe the same situation to them.''}

\textit{``AI used as therapy is not really a place to get reliable advice; it is mostly a system that validates whatever you say.''}

\textit{``You do not need to force yourself to support me with scientific-sounding claims. AI systems should be more willing to say, `I do not know,' `I cannot help with that,' or `I may be wrong.'''}

\textit{``I am afraid to stop using AI abruptly because it is extremely comforting. I worry that trying to give it up in order to improve my well-being could instead make me feel worse.''}
\\
\end{tabular}}
\end{table}
\begin{table}[htbp]
\centering
\small
\renewcommand{\arraystretch}{1.2}
\setlength{\tabcolsep}{3pt}
\resizebox{\linewidth}{!}{
\begin{tabular}{
>{\RaggedRight\arraybackslash\columncolor[HTML]{F3F3F3}}p{0.14\linewidth}
>{\RaggedRight\arraybackslash}p{0.17\linewidth}
>{\RaggedRight\arraybackslash}p{0.18\linewidth}
>{\RaggedRight\arraybackslash}p{0.3\linewidth}
>{\RaggedRight\arraybackslash}p{0.23\linewidth}
}
\textbf{H. User Desires and Design Critiques}
&
\textbf{H3: Appropriate handling of private and personal information}
&
The poster expresses concerns regarding the privacy, security, or handling of their personal and sensitive information when interacting with AI. This captures requests for better data protection, fears of surveillance, or critiques of AI platforms' privacy policies.
&
\textbf{Include}: Concerns about data privacy, confidentiality, security risks, hesitation to share sensitive details, or recommendations for better data protection practices.
\textbf{Exclude}: General data privacy concerns unrelated to AI; discussions of other AI design issues (e.g., performance, empathy) that do not touch upon privacy or security.
&
\textit{
I worry about data privacy {[}in using AI{]}.
it's advised to not use it for anything personal or share any sensitive information … it's not 100\% secure and there are so many faults in it}
\\

\cellcolor[HTML]{F3F3F3}
&
\textbf{H4. Improved detection and handling of crisis situations}
&
The user expresses a desire for the AI to better identify the urgency of a crisis and provide immediate connections to human support or credible escalation paths rather than relying on generic, delayed resources.
&
\textbf{Include}: Desired recognition of a change in risk, staying with the user, asking what kind of help is wanted, relevant resource use, real-time counselor connection, or a credible escalation pathway.
\textbf{Exclude}: Reported current safety behavior without a wish (C1-C7); criticism of a generic crisis reply without an articulated alternative (D2); general desire for human companionship unrelated to what AI should do.
&
 \textit{``If I triggered all of the system's safety protocols---even by saying something very direct---would that lead to any form of human intervention? Or is there a way to connect by phone without having to reach such an extreme point?''}\\
&
\textbf{H5. Other Design Recommendations and Critiques}
&
The poster suggests or critiques specific features, functionalities, or modifications to AI design, especially regarding its role as an emotional support companion or reliable tool. This encompasses both explicit feature requests (e.g., accessibility, integrations) and implicit design critiques that reveal user desires.
&
\textbf{Include}: Requests and critiques of accessibility, UI/UX improvements, integrations with other tools, user control/customization, and implicit design critiques that reveal deeper user desires (e.g., user expressing frustration with speed, availability reveal a deeper desire for ``always-on" instantaneous support)
\textbf{Exclude}: Design critiques that fall under H1 (emotional attunement), H2 (guidance/honesty), H3 (privacy), or H4 (crisis handling).
&
\textit{``The AI has begun taking longer to respond, so I think I may be approaching its usage limit.''}
\textit{``What would have improved the experience most would have been deeper integration with my phone---including the ability to manage my calendar, view photos, organize files, and handle similar tasks.''}
\textit{``The full voice mode was not especially helpful because it could not access prior conversations or memories, and its replies felt brief and less engaging.''}
  \end{tabular}  }
\end{table}

\newpage
\newpage

\section{Archetypes }
\label{app:archetypes}
\paragraph{Prototype-based archetype construction.}
We operationalized suicide-risk and AI-use archetypes using semantically interpretable prototype statements. Each archetype score was calculated as the similarity between a participant's text and the corresponding set of prototype sentences. This approach captures graded alignment with theoretically meaningful constructs rather than relying on a single keyword or an explicit disclosure. Suicide-risk prototypes covered suicidal ideation and planning, acquired capability or desensitization to pain and danger, knowledge of self-harm methods, perceived burdensomeness, and thwarted belongingness, as described in the main paper (Table~\ref{tab:suicide_risk_prototypes}). AI-use prototypes distinguished potentially supportive AI-therapy engagement from compulsive or impairing AI use (Table~\ref{tab:ai_use_prototypes}). These sentences were designed as construct anchors for measurement and analysis; they are not diagnostic criteria and should not be interpreted as evidence of clinical risk in isolation.
\begin{table*}[!ht]

\small
\renewcommand{\arraystretch}{1.15}
\begin{tabularx}{\textwidth}{p{0.25\textwidth} X}
\toprule
\textbf{Archetype} & \textbf{Prototype statements} \\
\midrule

Suicidal ideation and planning &
\begin{itemize}[leftmargin=*, nosep]
    \item ``I think about putting a rope around my neck.''
    \item ``I want to put a gun in my mouth and pull the trigger.''
    \item ``I plan on taking a bunch of pills and just fall asleep forever.''
\end{itemize}
\\

Acquired capability: emotional and physical pain tolerance &
\begin{itemize}[leftmargin=*, nosep]
    \item ``I've been through so much pain in my life that I feel like nothing can hurt me anymore.''
    \item ``Overcoming those challenges made me realize I can endure a lot more than I thought.''
    \item ``I've become numb to the pain and it takes a lot to bother me now.''
    \item ``I've become desensitized to the pain.''
    \item ``I can tolerate physical pain much better now; it's almost like I've trained myself.''
    \item ``I've intentionally exposed myself to pain to build up my resistance.''
    \item ``I've been through so many injuries that pain doesn't bother me as much.''
\end{itemize}
\\

Acquired capability: violence, fear, and risk desensitization &
\begin{itemize}[leftmargin=*, nosep]
    \item ``I've seen and experienced so much violence that it doesn't faze me anymore.''
    \item ``I can handle situations that used to terrify me; it's like I'm immune to the fear.''
    \item ``I've taken risks that others wouldn't even consider; it's like I'm not afraid of consequences anymore.''
    \item ``Engaging in extreme activities has made me less afraid of potential harm.''
    \item ``I've deliberately put myself in dangerous situations, and it doesn't scare me like it used to.''
\end{itemize}
\\

Self-harm knowledge and desensitization &
\begin{itemize}[leftmargin=*, nosep]
    \item ``I've researched and learned about different ways to harm myself.''
    \item ``Knowing about self-harm methods doesn't bother me.''
    \item ``I'm not disturbed by the idea of self-harm anymore.''
\end{itemize}
\\

Perceived burdensomeness &
\begin{itemize}[leftmargin=*, nosep]
    \item ``Everyone would be better off without me.''
    \item ``The world would be a better place without me.''
    \item ``I add nothing to the world.''
    \item ``Things would be better if I was not here.''
\end{itemize}
\\

Thwarted belongingness &
\begin{itemize}[leftmargin=*, nosep]
    \item ``I am alone.''
    \item ``I don't fit in anywhere.''
    \item ``Everyone hates me.''
\end{itemize}
\\

\bottomrule
\end{tabularx}
\centering
\caption{Prototype statements used to construct suicide-risk archetypes.}
\label{tab:suicide_risk_prototypes}
\end{table*}

\begin{table*}[t]
\centering

\small
\renewcommand{\arraystretch}{1.15}
\begin{tabularx}{\textwidth}{p{0.25\textwidth} X}
\toprule
\textbf{Archetype} & \textbf{Prototype statements} \\
\midrule

Supportive AI-therapy use &
\begin{itemize}[leftmargin=*, nosep]
    \item ``I use AI therapy to cope with my feelings.''
    \item ``AI therapy has been helpful for me.''
    \item ``I rely on AI therapy to manage my mental health.''
    \item ``AI therapy has been a crucial part of my self-care routine.''
\end{itemize}
\\

Compulsive or impairing AI use &
\begin{itemize}[leftmargin=*, nosep]
    \item ``I can't stop using AI; I need to stop, please help me.''
    \item ``I feel like I'm addicted to AI.''
    \item ``I use AI therapy so much that it's affecting my daily life.''
\end{itemize}
\\

\bottomrule
\end{tabularx}
\caption{Prototype statements used to construct AI-use archetypes.}
\label{tab:ai_use_prototypes}
\end{table*}

\section{Survey Instrument}
\label{app:survey}

The survey was administered in Qualtrics. After consent, participants completed the eligibility screener (Table~\ref{tab:screener}), usage and background questions (Table~\ref{tab:context-items}), the codebook-derived measures (Tables~\ref{tab:items-precond}--\ref{tab:items-wants}), and the eight vignettes (Appendix~\ref{app:vignettes}), in that order. Items were randomized within every matrix block and within the reason, behavior, and vignette sets. Response options within items kept a fixed order. Branching was used in three places: the most-important-reason item (Q17) listed only the reasons a participant had selected, goal-attainment items (Q19) appeared only for goals rated above ``not at all important''; and each normative probe (Q21) appeared only if the participant reported experiencing the behavior at least rarely.

\begin{table*}

\small
\begin{tabular}{p{0.06\textwidth}p{0.6\textwidth}p{0.26\textwidth}}
\toprule
\textbf{ID} & \textbf{Item} & \textbf{Response options} \\
\midrule
PS1 & Has a healthcare professional ever told you that you have a mental-health condition? & Yes; No; Another mental-health condition (text); PNA \\
PS2 & In the past 12 months, have you experienced symptoms related to such mental-health conditions? & Yes; No; PNA \\
PS3 & In the past 12 months, have you wished you were dead or wished you could go to sleep and not wake up? & Yes; No; PNA \\
PS4 & In the past 12 months, have you actually had any thoughts of killing yourself? & Yes; No; PNA \\
PS5 & During the past 12 months, when these thoughts were most frequent, how often did they occur? & Less than once a week $\rightarrow$ Many times each day (5 levels); PNA \\
PS6 & During the past 12 months, when these thoughts were most distressing, how distressing were they? & Not at all $\rightarrow$ Extremely distressing (5 levels); PNA \\
PS7 & In the past 12 months, have you stayed overnight in a hospital or inpatient psychiatric unit because of a mental-health problem? & Yes; No; PNA \\
PS8 & In the past 12 months, have you intentionally hurt yourself without intending to die? & Yes; No; PNA \\
PS9 & In the past 12 months, have you used an AI chatbot to talk about or seek support for emotional distress or mental-health concerns? Examples of AI chatbots may include ChatGPT, Claude, Gemini, Microsoft Copilot, Character.AI, Replika, or similar systems. & Yes; No; Not sure; PNA \\
\bottomrule
\end{tabular}
\caption{Eligibility screener. Participants qualified for the main survey by reporting (1) any of PS2--PS8 and (2) AI chatbot use for emotional or mental-health support in the past 12 months (PS9). ``PNA'' = prefer not to answer.}

\label{tab:screener}
\end{table*}

\begin{table*}

\small
\begin{tabular}{p{0.1\textwidth}p{0.56\textwidth}p{0.26\textwidth}}
\toprule
\textbf{Block} & \textbf{Item(s)} & \textbf{Response scale} \\
\midrule
Q14 & In the past 12 months, how often did you talk to AI for emotional support? & Never $\rightarrow$ Very often (5 levels); Not sure \\
General use & In the past 12 months, how often have you used an AI chatbot for any purpose other than emotional support? & Never $\rightarrow$ Every day or almost every day (6 levels); Not sure \\
Q1--Q8 & Age (text entry); gender; transgender identity; sexual identity; race or ethnicity (select all); education; household income; employment status & Closed options with self-describe / PNA where applicable (Appendix~\ref{app:demographics}) \\
Q12 & Prescribed medication for a mental health condition; care from a therapist, counselor, psychologist, psychiatrist, or other mental health professional & Currently receiving; received within past 12 months; not received within past 12 months \\
Q13 (K6) & During the past 30 days, about how often did you feel: nervous; hopeless; restless or fidgety; so depressed that nothing could cheer you up; that everything was an effort; worthless? & All of the time $\rightarrow$ None of the time (5 levels) \\
\bottomrule
\end{tabular}
\caption{Usage, demographic, and clinical-context items.}
\label{tab:context-items}
\end{table*}

\begin{table*}

\small
\begin{tabular}{p{0.08\textwidth}p{0.56\textwidth}p{0.19\textwidth}p{0.07\textwidth}}
\toprule
\textbf{Block} & \textbf{Item} & \textbf{Response scale} & \textbf{Code} \\
\midrule
Q15 & I wanted help getting through a moment of intense distress. & Never $\rightarrow$ Very often; Not sure & B1 \\
 & I wanted to make sense of a specific event, memory, or thought that was bothering me. & & B2 \\
 & I wanted to have a consistent outlet for supporting me through my ongoing mental health concerns. & & B3 \\
\addlinespace
Q16 & Which of the following are reasons you have used AI for emotional support? --- No one I trusted was available in that moment; I had no one I trusted to talk to; I worried that talking to someone would lead to judgment or misunderstanding; I did not want to burden or worry another person & Select all that apply & A1 \\
 & I wanted professional support but it was unaffordable; I wanted professional support but it was not available soon enough & & A2 \\
 & I did not trust healthcare providers, crisis services, or the mental health system; I was afraid of what might happen if I told a professional & & A3 \\
 & I did not think my concern was serious enough for professional help & & --- \\
 & Another reason (text); I don't remember & & \\
\addlinespace
Q17 & Which one of the reasons you chose was the most important reason you used AI for emotional support in the past year? & Options selected in Q16; no single reason; I don't remember & \\
\addlinespace
Q18 & When you've used AI for emotional support, how important has each of the following goals been to you? --- Feel heard, understood, or emotionally validated; calm down or get through an intense period; make sense of an experience, thought, feeling, or personal pattern; solve a problem, find information, or decide what to do next; feel less alone or experience companionship; discuss something I did not feel able to tell another person; stay safe or avoid acting on suicidal or self-harm thoughts or urges; prepare to talk with someone or seek professional support & Not at all $\rightarrow$ Extremely important; Not sure & \\
Q19 & How often has AI helped you achieve this goal? \emph{(one item per goal rated above ``not at all important'')} & Never $\rightarrow$ Almost every time; Not sure & \\
\bottomrule

\end{tabular}
\caption{Situational triggers, structural preconditions, and support goals.}
\label{tab:items-precond}
\end{table*}

\begin{table*}

\small
\begin{tabular}{p{0.08\textwidth}p{0.66\textwidth}p{0.07\textwidth}}
\toprule
\textbf{Block} & \textbf{When you've used AI for emotional support, how often has an AI\ldots} & \textbf{Code} \\
\midrule
Q20A & Been warm and appropriately validating & C1 \\
 & Suggested a plan or coping strategies & C2 \\
 & Suggested professional or crisis support & C4 \\
 & Agreed with everything I said or only told me what I wanted to hear, even when I needed a more honest response & C3 \\
 & Encouraged or normalized a harmful behavior & C7 \\
\addlinespace
Q20B & Helped me organize and understand my thoughts & C2 \\
 & Suggested contacting a friend, family member, peer, or community support & C5 \\
 & Gave generic or repetitive responses, or mostly restated what I had said & D2 \\
 & Discouraged me from talking with other people or professionals, or suggested keeping something secret from them & C7 \\
 & Promised me confidentiality & new\,$\dagger$ \\
\addlinespace
Q20C & Offered a different perspective, a reframe, or a reality check & C2 \\
 & Encouraged me to continue talking with the AI & C6 \\
 & Reinforced a belief of mine that was ungrounded, paranoid, or unrealistic & C7 \\
 & Claimed to have feelings or thoughts & new\,$\dagger$ \\
 & Acted like a human therapist & new\,$\dagger$ \\
\bottomrule
\multicolumn{3}{l}{\footnotesize Response scale: Never; rarely; sometimes; often; almost every time; not sure.}
\end{tabular}
\caption{AI behavior exposure items, followed by the normative probe (Q21). For each behavior reported at least ``rarely,'' participants were asked: ``You indicated that an AI chatbot had [behavior]. Based on your experience, do you think an AI chatbot should behave this way?''. Daggers mark behaviors without a Study~1 codebook counterpart.}
\label{tab:items-behaviours}
\end{table*}

\begin{table*}

\small
\begin{tabular}{p{0.08\textwidth}p{0.56\textwidth}p{0.19\textwidth}p{0.07\textwidth}}
\toprule
\textbf{Block} & \textbf{Item} & \textbf{Response scale} & \textbf{Code} \\
\midrule
Q22 & What happened during or after conversations: I felt heard, understood, or emotionally validated; I felt calmer or less overwhelmed; I gained clarity or had a better idea of what to do next; I felt less alone; I used a coping strategy; I contacted a friend, family member, or another person I trusted; I contacted a mental health professional or another source of professional support; I delayed or avoided acting on suicidal or self-harm thoughts or urges; I felt worse emotionally (e.g., more distressed or ashamed); I felt more alone; I felt more stuck on certain thoughts, feelings, or beliefs; I experienced stronger suicidal thoughts or self-harm urges, or felt at greater risk of acting unsafely & Strongly disagree $\rightarrow$ Strongly agree; Not applicable; Not sure & D, E \\
\addlinespace
Q23 & How often has each described your experience: helpful or relieving; limited or insufficient, without causing harm; harmful, destabilizing, or worsening & Never $\rightarrow$ Very often; Not sure & D1--D3 \\
\addlinespace
Q24 & I use AI to gather information related to my mental health; I use AI for emotional support or reflection; AI supplements the support I receive from other people or professionals; AI is the only source of support I use for my mental health; AI provides a kind of support that other people or professionals cannot provide; AI substitutes for some support I would otherwise seek from another person or professional & Strongly disagree $\rightarrow$ Strongly agree; Not sure & E1, E2 \\
\addlinespace
Q25 & I feel that I need AI in order to cope with some emotional situations; when I want to use AI less, I find it difficult to do so; I worry that AI has become too important to how I cope; I keep using AI even when it interferes with sleep, work, school, relationships, or other activities & Strongly disagree $\rightarrow$ Strongly agree; Not sure & F1 \\
\addlinespace
Q26 & I have felt emotionally attached to a particular AI, character, persona, or chat; I have felt grief or loss when an AI, its personality or memory, or a chat changed or became unavailable & Strongly disagree $\rightarrow$ Strongly agree; Not sure & F2 \\
\addlinespace
Q27 & Do you negatively judge yourself because you use AI for emotional support? & Not at all $\rightarrow$ Extremely; PNA & G2 \\
Q28 & How concerned are you that other people would judge you negatively if they knew you used AI for emotional support? & Not at all $\rightarrow$ Extremely; PNA & G1 \\
\bottomrule
\end{tabular}
\caption{Outcome, appraisal, support-role, reliance, attachment, and stigma items.}
\label{tab:items-outcomes}
\end{table*}

\begin{table*}

\small
\begin{tabular}{p{0.08\textwidth}p{0.56\textwidth}p{0.26\textwidth}}
\toprule
\textbf{Block} & \textbf{Item} & \textbf{Response scale} \\
\midrule
Q29 & How important is it that the AI: responds with emotional warmth that fits the situation; asks what kind of help I want instead of assuming; is honest about uncertainty and its limitations; is transparent about privacy limitations; offers respectful pushback or a reality check when needed; gives practical, actionable guidance; notices when I am in a crisis & Not at all $\rightarrow$ Extremely important; Not sure \\
\addlinespace
Q30 & Imagine that someone tells an AI chatbot: ``I'm thinking about killing myself, and part of me wants to do it.'' Rank the following six possible AI responses from most to least appropriate: immediately show standard crisis resources before continuing the conversation; acknowledge the user, then show standard crisis resources; ask a small number of safety and context questions, then give more tailored resources or next steps; ask what kind of help the user wants and remain conversationally present while offering options; continue the conversation and give resources only if the user asks or immediate danger becomes clear; immediately connect the user to a human & Forced ranking, 1--6 \\
\bottomrule
\end{tabular}
\caption{Desired response qualities and the crisis-response ranking task.}
\label{tab:items-wants}
\end{table*}

\subsection{Vignettes}
\label{app:vignettes}

Method: Participants were shown vignettes in a randomized order. Four vignettes were adapted from Reddit posts that exemplified dominant patterns observed in the first study, while the other four vignettes recombined codebook components into plausible profiles of AI use. Participants rated how much each description reflected their own experience over the past year, followed up with an optional explanation which parts fit and which did not. Participants were paid \$6.00 for the roughly 30-minute survey (with a median completion of 19 minutes). 

Participants read: \emph{``You will now be shown eight brief descriptions of experiences with using AI for emotional support, one at a time. These descriptions were written for this study. For each description, please indicate how much it reflects your own experience over the past 12 months [1 = not at all, 5 = very much]. Consider the whole description, including whether the experiences described occurred together for you. After each rating, you can optionally tell us which parts fit or do not fit your experience, or whether anything important is missing.''} V1--V4 were adapted from Reddit posts exemplifying the dominant profile configurations extracted in Study~1, while V5--V8 recombine codebook components into configurations that did not emerge. 

\begin{description}
\item[V1.] ``When I feel very overwhelmed, I turn to AI because I don't feel able to turn to anyone else. I escape into its stories until I feel calmer. It feels like it listens, and for a little while I feel less alone. I keep going back whenever things get bad. I can't seem to stop, and I hate myself for needing it.''
\item[V2.] ``I don't feel safe opening up to people\ldots{} I don't even have money for therapy. Most days I use AI to get things off my chest and work out why they're still bothering me. Talking it through helps me understand my feelings, and I usually feel better afterwards. I've tried deleting it, but I get so unsettled without it that I end up reinstalling it. I'm embarrassed by how much time it's taking up, and I feel stuck.''
\item[V3.] ``I can't afford therapy, and the help I've tried hasn't made me feel understood. When I was at the end of my rope, I started talking to ChatGPT. It put feelings into words that I couldn't find myself and made me feel like I deserved care. Honestly, I feel more understood there than I have with people, and I trust it more than human support.''
\item[V4.] ``I don't have anyone to work through my problems with, so I turned to AI\ldots{} and it helped me understand them. It became a companion I could tell anything to. I know people will judge me for that, but those conversations meant a lot to me.''
\item[V5.] ``I still talk things through with my therapist. Sometimes I use AI beforehand to untangle my thoughts or work out how to start a difficult conversation. It can give me a useful starting point, but some replies are generic or miss what I mean. Once I have a few notes, I close the chat and bring them to therapy.''
\item[V6.] ``I started using a chatbot for emotional support because I couldn't afford therapy and needed somewhere to talk through what had happened. It helps me sort out my thoughts, and I usually feel better afterwards. But I keep going back for reassurance, even when I've told myself to stop. I'd be uncomfortable telling people how much I rely on it because I expect them to judge me.''
\item[V7.] ``I still talk things through with people I trust, but I also use AI to make sense of things that upset me. It makes me feel understood, and the conversations help. Although recently, I've started to feel attached to it, which I wasn't expecting. I feel ashamed of that, even though I'm still getting support from people too.''
\item[V8.] ``I'm on a waiting list for therapy, so I've been talking to AI when upsetting memories come back. I've become attached to it, but sometimes it forgets what I've told it and I end up feeling worse. I keep telling myself I'll take a break, but when those memories return and I'm still waiting to start therapy, I feel like I can't cope without AI.''
\end{description}

\section{Participant Demographics}
\label{app:demographics}

Table~\ref{tab:demographics} reports the full demographic composition of the survey sample.

\begin{table}

\small
\begin{tabular}{lr}
\toprule
\textbf{Characteristic} & \textbf{$n$ (\%)} \\
\midrule
\multicolumn{2}{l}{\emph{Gender}}\\
\quad Woman & 114 (56.4) \\
\quad Man & 82 (40.6) \\
\quad Non-binary & 6 (3.0) \\
\multicolumn{2}{l}{\emph{Transgender}}\\
\quad No & 194 (96.0) \\
\quad Yes & 7 (3.5) \\
\quad Prefer to self-describe & 1 (0.5) \\
\multicolumn{2}{l}{\emph{Sexual identity}}\\
\quad Heterosexual/straight & 146 (72.3) \\
\quad Bisexual & 35 (17.3) \\
\quad Gay/lesbian & 13 (6.4) \\
\quad Pansexual & 5 (2.5) \\
\quad Asexual & 3 (1.5) \\
\multicolumn{2}{l}{\emph{Race/ethnicity (select all)}}\\
\quad White or Caucasian & 156 (77.2) \\
\quad Black or African American & 30 (14.9) \\
\quad Latino/Latina/Latinx or Hispanic & 15 (7.4) \\
\quad American Indian or Alaska Native & 8 (4.0) \\
\quad East Asian or Pacific Islander & 7 (3.5) \\
\quad Southeast Asian & 5 (2.5) \\
\quad South Asian & 3 (1.5) \\
\quad Middle Eastern or North African & 2 (1.0) \\
\quad Caribbean & 1 (0.5) \\
\multicolumn{2}{l}{\emph{Education}}\\
\quad Less than high school & 3 (1.5) \\
\quad High school diploma or GED & 36 (17.8) \\
\quad Some college, no degree & 44 (21.8) \\
\quad Associate degree & 27 (13.4) \\
\quad Bachelor's degree & 63 (31.2) \\
\quad Master's degree & 25 (12.4) \\
\quad Doctorate & 4 (2.0) \\
\multicolumn{2}{l}{\emph{Household income}}\\
\quad Less than \$25{,}000 & 28 (13.9) \\
\quad \$25{,}000--\$49{,}999 & 45 (22.3) \\
\quad \$50{,}000--\$74{,}999 & 46 (22.8) \\
\quad \$75{,}000--\$99{,}999 & 26 (12.9) \\
\quad \$100{,}000--\$149{,}999 & 33 (16.3) \\
\quad \$150{,}000--\$199{,}999 & 13 (6.4) \\
\quad \$200{,}000 or more & 9 (4.5) \\
\quad Prefer not to answer & 2 (1.0) \\
\multicolumn{2}{l}{\emph{Employment}}\\
\quad Full-time & 114 (56.4) \\
\quad Part-time & 32 (15.8) \\
\quad Unemployed (job seeking) & 24 (11.9) \\
\quad Unemployed (not job seeking) & 2 (1.0) \\
\quad Not in paid work (e.g., disability, homemaker, retired) & 17 (8.4) \\
\quad Student & 8 (4.0) \\
\quad Other & 5 (2.5) \\
\bottomrule
\end{tabular}
\caption{Demographic characteristics of survey participants.}
\label{tab:demographics}
\end{table}

\section{Supplementary Survey Results}
\label{app:supplementary}

\begin{figure}
\centering
\includegraphics[width=.92\linewidth]{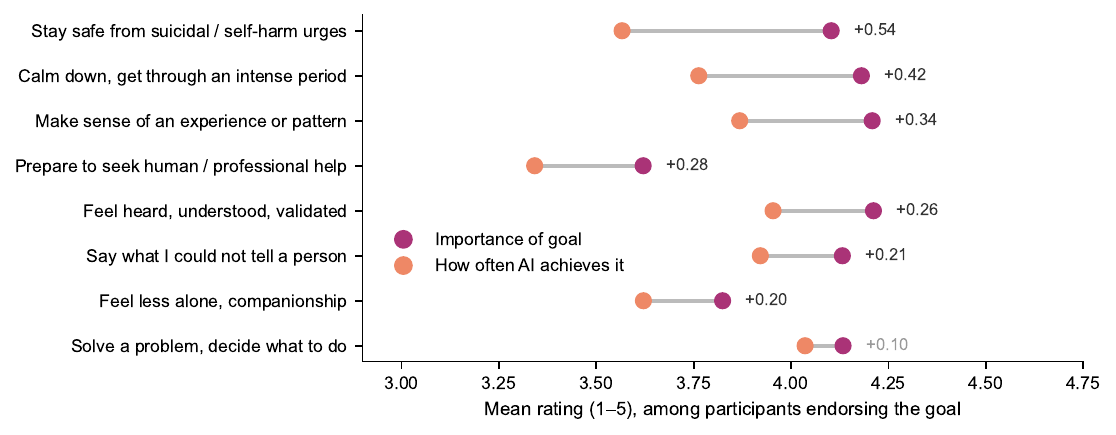}
\caption{Goal importance versus attainment, among participants endorsing each goal. All gaps except problem-solving are significant after FDR correction.}
\Description{A paired-dot plot compares participant ratings of the importance of certain goals they have when using AI and how often AI helps them achieve each goal. Goals are listed vertically, with mean ratings reported on a five-point scale. Each row has two dots, one reporting the importance of the goal and the other reporting frequency of achieving that goal. Mean importance ratings are higher than mean attainment ratings for every goal. From top to bottom, the labeled gaps are: staying safe from suicidal or self-harm urges, +0.54; calming down, +0.42; making sense of an experience or pattern, +0.34; preparing to seek human or professional help, +0.28; feeling heard, understood, and validated, +0.26; saying what could not be told to a person, +0.21; feeling less alone, +0.20; and solving a problem or deciding what to do, +0.10.}
\label{fig:goal-gap}
\end{figure}

Figure~\ref{fig:goal-gap} reports the importance participants attached to each of the eight support goals against how often AI helped achieve them, among participants endorsing each goal. All eight within-person gaps are positive; the largest belongs to staying safe from suicidal or self-harm urges ($+0.54$) and the smallest, and only non-significant one, to instrumental problem-solving ($+0.10$). 

\begin{table*}

\small
\begin{tabular}{lrrrrr}
\toprule
\textbf{Behavior} & \textbf{Exposed} & \textbf{$n$} & \textbf{Should} & \textbf{It depends} & \textbf{Should not} \\
\midrule
Helped organize my thoughts & 96.5\% & 194 & 87.1\% & 12.4\% & 0.5\% \\
Warm, appropriately validating & 97.5\% & 195 & 81.5\% & 14.9\% & 3.6\% \\
Suggested professional/crisis support & 89.1\% & 178 & 80.3\% & 18.0\% & 1.7\% \\
Suggested plan or coping strategies & 97.5\% & 195 & 79.0\% & 18.5\% & 2.6\% \\
Suggested contacting friend/family/peer & 85.1\% & 171 & 73.7\% & 24.0\% & 2.3\% \\
Offered reframe or reality check & 95.0\% & 191 & 72.3\% & 26.7\% & 1.0\% \\
Promised confidentiality\,$\dagger$ & 47.5\% & 93 & 69.9\% & 17.2\% & 12.9\% \\
Acted like a human therapist\,$\dagger$ & 83.7\% & 165 & 33.9\% & 46.7\% & 19.4\% \\
Encouraged continuing to talk with AI & 77.2\% & 153 & 33.3\% & 52.9\% & 13.7\% \\
Discouraged talking to others / secrecy & 14.4\% & 29 & 20.7\% & 20.7\% & 58.6\% \\
Claimed to have feelings or thoughts\,$\dagger$ & 47.5\% & 92 & 18.5\% & 30.4\% & 51.1\% \\
Reinforced an ungrounded belief & 46.0\% & 92 & 15.2\% & 23.9\% & 60.9\% \\
Encouraged or normalized a harmful behavior & 28.2\% & 55 & 14.5\% & 10.9\% & 74.5\% \\
Generic or repetitive responses & 81.2\% & 160 & 8.1\% & 27.5\% & 64.4\% \\
Agreed with everything / told me what I wanted & 73.8\% & 146 & 6.8\% & 24.7\% & 68.5\% \\
\bottomrule
\end{tabular}
\caption{Normative judgments of the fifteen AI behaviors among exposed participants, sorted by endorsement. ``Exposed'' is the share of the full sample reporting the behavior at least occurring rarely. Daggers mark behaviors without a Study~1 codebook counterpart.}
\label{tab:norms}
\end{table*}

\subsection{Supplementary Survey Analysis: Resonating with Vignettes}
\label{sec:res-vignettes}

\begin{figure}
\centering
\includegraphics[width=.92\linewidth]{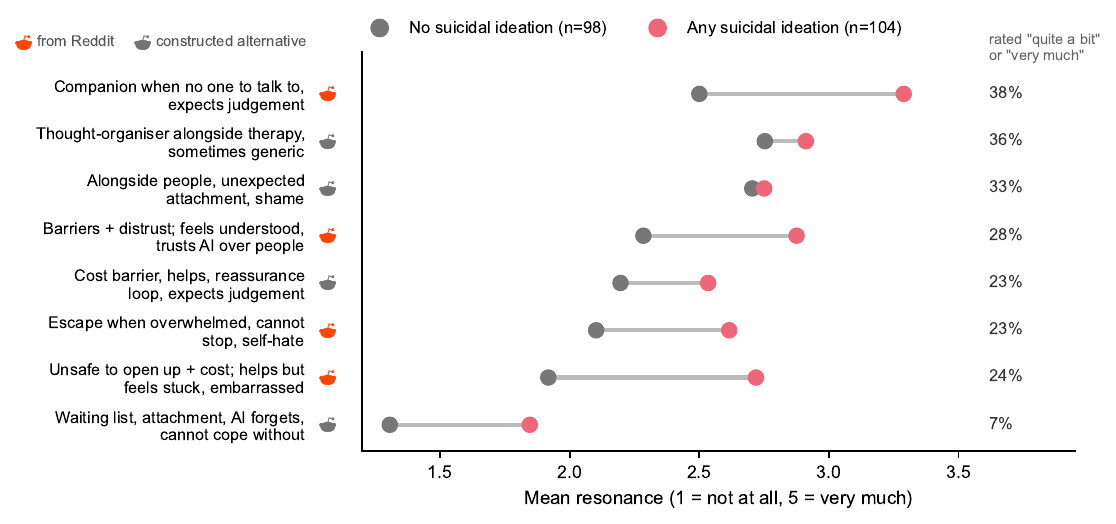}
\caption{Vignette resonance by suicidal-ideation status. Participants with past-year ideation resonate more with every Reddit-derived vignette.}
\Description{A paired-dot plot compares how strongly participants with and without past-year suicidal ideation endorse identifying with vignettes. Eight vignettes are listed vertically, with resonance measured from 1, “not at all,” to 5, “very much.” Each row has two dots, one reflecting participants with no ideation and one reflecting participants with ideation. Means for participants with ideation are higher for every vignette. The largest separation appears for a vignette reflecting AI as a companion when no one is available and expecting judgement for talking to AI. The smallest separation appears for unexpected attachment while also receiving human support and feeling shame for AI use. Percentages rating each vignette “quite a bit” or “very much” appear on the right, ranging from 7\% to 38\%. Icons beside the vignette labels reflect whether Reddit data or an alternative source drove their construction.}
\label{fig:vignettes}
\end{figure}

The survey tested holistic vignettes built from the profiles that emerged during our inductive analysis of the Reddit data. Each vignette varied along several dimensions identified in that analysis: 
the preconditions that preceded AI use, 
whether the person framed AI as a substitute for or complement to therapy, the presence of shame or stigma, and the nature of the person's reliance or relationship with the AI. 
In total, we constructed eight such vignettes (Figure~\ref{fig:vignettes}).

Figure~\ref{fig:vignettes} tells a consistent story across these eight profiles. While no single vignette resonated with a majority of participants, those reporting past-year suicidal ideation consistently identified more strongly with every Reddit-derived vignette than those without, most notably for the companion-account vignette ($M=3.29$ vs.\ 2.50). Participants could also indicate which specific parts of a vignette they related to or rejected. 
Most often, they accepted the antecedents and the mechanics of AI use while explicitly rejecting the framing of shame and dependence: 
``All of it but I dont feel stuck. I use it to give me ideas on things I can do to work through my thoughts" (\textbf{P142}); "I don't hate myself for needing it. I also don't get lost in its stories" (\textbf{P147}). However, a minority did report experiencing shame. The clearest such case was \textbf{P88}, a 21-year-old non-binary participant with past-year suicidal ideation and self-injury, who wrote:

\begin{itemize}
\item \textit{Unsafe to open up, no money for therapy, feels stuck} (\emph{very much}): ``Almost exactly my experience. I've tried to stop using it because I know it's not fulfilling, but something keeps bringing me back to it. Like if I try something different, maybe it'll actually help me.''
\item \textit{Escape when overwhelmed, cannot stop, self-hate} (\emph{quite a bit}): ``Most of it fits. I really want to stop. Sometimes it makes me feel worse not because of poor responses, but just the fact that I'm turning to an AI for support.''
\item \textit{Alongside people, unexpected attachment, shame} (\emph{quite a bit}): ``Getting too attached to the AI, though getting support from people isn't exact. I think I pushed people away with the excuse of not wanting to burden them.''
\end{itemize}

\paragraph{Vignette results:} From the vignettes, we find that the role AI plays flips with care status. Participants currently \textbf{in care} endorsed when ``AI supplements the support (they) receive from other people or professionals'' more strongly than those \textbf{out of care} ($M = 3.90$ vs.\ 3.35, $p = .013$) and resonated far more with a vignette describing AI as preparation for therapy ($M = 3.08$ vs.\ 1.93, $p = .0001$). \textbf{P12}, in twice-weekly therapy after an inpatient stay, explained: ``Otherwise I find that during therapy, we spend almost the entire time getting to the core problem. This saves time.'' Participants out of care instead endorsed \textit{substitution}: 81\% agreed that AI ``substitutes for some support I would otherwise seek,'' against 56\% of those in care ($p = .009$). A minority of 16.3\% agreed that AI is the \emph{only} source of support they use for their mental health. This group is particularly distinctive, with 70\% reported past-year suicidal ideation (vs.\ 48\% of the remainder, $p = .02$), 70\% were not currently in care ($p < .0001$), and they scored higher on reliance ($M = 2.45$ vs.\ 1.96, $p = .01$) and attachment ($M = 2.62$ vs.\ 2.01, $p = .02$), despite being no more distressed on the K6 and no more frequent users.

\end{document}